\documentclass[11pt,a4paper]{article}
\pdfoutput=1
\usepackage{jheppub}
\usepackage[T1]{fontenc}
\usepackage[spanish,USenglish]{babel}
\usepackage{braket}
\usepackage{bm}
\usepackage{microtype}
\usepackage{booktabs}

\DeclareMathOperator{\diag}{diag}

\newcommand{\smallcg}[6]{
\Braket{
\begin{smallmatrix}
#1	& #3	\\
#2	& #4	\\
\end{smallmatrix}
|
\begin{smallmatrix}
#5	\\
#6 	\\
\end{smallmatrix}
}
}
\newcommand{\cgr}[6]{
\Biggl\langle
\begin{array}{@{}cc}
#1	& #3		\\
#2	& #4	\\
\end{array}
\Bigg\vert
\begin{array}{c@{}}
#5	\\
#6 	\\
\end{array}
\Biggr\rangle
}
\newcommand{\cgl}[6]{
\Biggl\langle
\begin{array}{@{}c}
#1	\\
#2	\\
\end{array}
\Bigg\vert
\begin{array}{cc@{}}
#3	& #5	\\
#4	& #6 	\\
\end{array}
\Biggr\rangle
}

\begin{document}

\title{Heavy-quark spin symmetry breaking\texorpdfstring{\\}{ }in the Born-Oppenheimer approximation}
\author{R. Bruschini}
\affiliation{Department of Physics, The Ohio State University, Columbus, OH 43210, USA}
\emailAdd{bruschini.1@osu.edu}
\keywords{Properties of Hadrons, Quarkonium}
\arxivnumber{2303.17533}

\abstract{Heavy-quark spin symmetry is explicitly broken by the mass splitting between a heavy-light pseudoscalar meson and its vector partner. This fact plays a pivotal role in the physics of states whose mass lies close to the threshold of an open-flavor meson pair, like $X(3872)$. We show that this source of heavy-quark spin symmetry breaking can be systematically included within the diabatic representation of the Born-Oppenheimer approximation. We verify that including all the appropriate coupled channels guarantees conservation of total angular momentum, parity, and charge-conjugation parity. This marks a fundamental step towards a unified, first-principles study of quarkonia and meson molecules with hidden flavor.}

\maketitle

\section{Introduction}

In the Born-Oppenheimer (BO) approximation for QCD, the spectrum of a heavy quark-antiquark system is determined by potentials which are numerically accessible in lattice QCD \cite{Jug99}. Because of this feature, the BO picture is one of the most promising tools for the study of conventional and unconventional heavy quark-antiquark states from first principles; see, for instance, refs.~\cite{Braa14,Bra17} and references therein. The idea of using lattice QCD calculations of the BO potentials that include the effects of string breaking to precisely determine the spectrum and decays of heavy quark-antiquark systems has been around for quite some time \cite{Dru99}. Although direct lattice measurements of string breaking have been available for many years \cite{Bal05,Bul19}, achieving such a description remains a current theoretical challenge.

Until recently, the main difficulty consisted in overcoming the single-channel approximation traditionally associated with BO \cite{Braa14}, as string breaking couples quark-antiquark and di-meson channels. Significant theoretical progress in the application of BO with coupled channels has been reported. This includes, for instance, calculations of the bottomonium spectrum in BO using the first calculation of string breaking in lattice QCD \cite{Bic20,Bic21}, phenomenological studies of quarkoniumlike mesons and open-flavor di-meson scattering in the diabatic representation of BO \cite{Bru20,Bru21c}, analyses of exotic heavy hadrons in the diabatic representation of the diquark model \cite{Leb22}, and calculations of threshold effects in a BO effective field theory \cite{Tar22}.

Nevertheless, there is yet another constraint customarily linked with BO that must be removed in order to carry out precise, \textit{ab initio} studies of heavy quark-antiquark states: heavy-quark spin symmetry (HQSS). Current BO calculations of the bottomonium spectrum  with string breaking \cite{Bic20,Bic21} rely on HQSS, since it is realized for lattice QCD simulations with static quarks. However, because of the mass splitting between a heavy-light pseudoscalar meson $B$ and its vector partner $B^\ast$, HQSS is broken by the mixing of quarkoniumlike resonances with di-meson pairs. In fact, this HQSS-breaking effect has been used to explain the hadronic decays of $\Upsilon(4S)$ and $\Upsilon(10860)$; see, for instance, refs.~\cite{Vol12,Bar23} and references therein. In light of this, one could question the reliability of HQSS-based calculations of coupled quark-antiquark and di-meson channels, especially for energies close to a meson-pair threshold. The situation is even more problematic in the charmoniumlike sector, with a $D^\ast$-$D$ splitting three times bigger than the $B^\ast$-$B$ splitting. Phenomenological analyses \cite{Bru22} indicate that the precise location of the di-meson thresholds may be decisive for the understanding of exotic states like $X(3872)$.

The problem of HQSS breaking in the BO approximation has been addressed in the case of $ud\bar{b}\bar{b}$ tetraquarks by including the $B^\ast$-$B$ mass splitting and solving the resulting Schr\"odinger equations for the coupled BO channels \cite{Bic16}. However, the $ud\bar{b}\bar{b}$ tetraquark case is made somewhat simpler by the absence of any string breaking, hence the formalism described in ref.~\cite{Bic16} is not easily extended to other situations where string breaking plays a prominent role.

In this work, we show that HQSS breaking can be systematically and consistently included in the diabatic representation of the BO approximation. The effect of HQSS breaking is to split the thresholds and replace the single string-breaking transition rate from lattice QCD by a matrix of transition rates between quark-antiquark and di-meson configurations, with coefficients determined from Dirac algebra. The quark-antiquark static energy, di-meson thresholds, and the transition rates form an interaction matrix that depends on the spin configuration of the heavy sources. Linear combinations of the interaction matrix elements for different spin configurations, with coefficients determined from angular momentum theory, form the entries of a potential matrix from which the spectrum of the heavy quark-antiquark system can be calculated.  As a byproduct, we verify that the cylindrical symmetry group of a static quark-antiquark pair, $D_{\infty h}\otimes CP$, gets naturally promoted to the full $O(3)\otimes C$ symmetry group of QCD when the kinetic energies of the heavy quarks are introduced.

This result constitutes a substantial amendment to the previous phenomenological analyses in refs.~\cite{Bru20,Bru21c} by determining the matrix of coupling coefficients that in these studies was arbitrarily fixed \textit{ad hoc}. It provides a foothold for a detailed, unified study of heavy quark-antiquark systems from \textit{ab initio} lattice QCD simulations with static quarks.

The contents of this paper are organized as follows. In section~\ref{statsec}, we thoroughly review the static approximation for a heavy quark-antiquark pair. In section~\ref{stringsec}, we calculate the transition rates between quark-antiquark and the various di-meson configurations in terms of the string-breaking transition rate. In section~\ref{threshsec}, we include violation of HQSS by threshold spin splittings. In section~\ref{bosec}, we introduce the kinetic energies of the heavy quarks in the diabatic representation of BO and write down the master Schr\"odinger equation with quark-antiquark and di-meson channels. Finally, in section~\ref{oversec}, we summarize our findings and indicate possible future developments.

\section{Static approximation}
\label{statsec}

For the system of a heavy quark-antiquark pair with strong interactions, the time scale for the motion of the heavy quarks, dictated by the heavy-quark mass $m_Q$, is considerably larger than the time scale for the evolution of the gluon and light-quark fields, dictated by the nonperturbative QCD energy scale $\Lambda_\textup{QCD}$. Because of this sharp separation of the time scales involved, it makes sense to expand the Hamiltonian of the heavy quark-antiquark system in powers of $\Lambda_\textup{QCD}/m_Q$. The leading order contribution in this expansion corresponds to the infinite mass limit for the heavy quarks, $m_Q\to\infty$. In this limit, the heavy quark-antiquark pair has no motion and acts as a static source for the light QCD fields. The energy levels of the dynamical fields, that is, gluons and light quarks, are not affected by the spins of the static quarks. This decoupling of the heavy-quark spins is referred to as HQSS.

Hence, let us begin by considering QCD with static quark ($Q$) and antiquark ($\bar{Q}$) sources at the positions $+\frac{1}{2}\bm{r}$ and $-\frac{1}{2}\bm{r}$. In this static approximation, $\bm{r}$ is a constant parameter. The dynamical fields are gluons and light quarks ($q$) and antiquarks ($\bar{q}$). The relevant symmetry group of QCD is $O(3)\otimes C$, where $O(3)$ is the group formed by three-dimensional rotations and the parity transformation $P$, and $C$ is the charge conjugation. However, the static quarks break the symmetry down to the cylindrical subgroup $D_{\infty h}\otimes CP$, which consists of:
\begin{itemize}
\item rotations around the $\bm{\hat{r}}$ axis;
\item a reflection $\mathcal{R}$ through a plane containing $\bm{r}$;
\item the combined transformation $CP$.
\end{itemize}

There are several terms in the \emph{total angular momentum} $\bm{J}$ of a heavy quark-antiquark system. We denote the \emph{light $QCD$ angular momentum} of the dynamical fields by $\bm{J}_\textup{light}$. The \emph{total spin of the heavy quarks} is $\bm{S}_{Q\bar{Q}}$. The \emph{orbital angular momentum} of the sources is $\bm{L}$. The total angular momentum is then
\[
\bm{J} = \bm{L} + \bm{S}_{Q\bar{Q}} + \bm{J}_\textup{light}.
\]
It is also convenient to introduce the \emph{static spin} defined by
\[
\bm{S} = \bm{S}_{Q\bar{Q}} + \bm{J}_\textup{light}.
\]

The representations of $D_{\infty h}\otimes CP$ are conventionally labeled as $\Lambda_\eta^\epsilon$, where:
\begin{itemize}
\item $\Lambda= \lvert \bm{J}_\textup{light} \cdot \bm{\hat{r}} \rvert$ is the modulus of the projection of the light QCD angular momentum $\bm{J}_\textup{light}$ on $\bm{\hat{r}}$. The values of $\Lambda$ are indicated by Greek letters: $\Sigma$, $\Pi$, $\Delta$ for $\Lambda=0,1,2$, respectively, and so on.
\item $\eta=g$ or $u$ stands for $CP=+$ or $-$, respectively.
\item $\epsilon=\pm$ is the eigenvalue of $\mathcal{R}$ in the case $\Lambda=0$, and it is omitted for $\Lambda > 0$.
\end{itemize}
The energy levels of light QCD in the presence of the static sources can be labeled by the light BO quantum numbers $\Lambda_\eta^\epsilon$.

One could also define static BO quantum numbers for the system that consists of the light QCD fields and the spins of the heavy quarks, with static spin quantum number $s$ and projection $\lambda=\bm{S}\cdot\bm{\hat{r}}$.  In the case $\lambda=0$, the value of $\epsilon$ is completely determined by the static spin $s$, the intrinsic parity $\mathfrak{p}$ of the source, and the light BO quantum numbers $J_\textup{light}$ and $\epsilon_\textup{light}$ as $\epsilon=\epsilon_\textup{light}\mathfrak{p}(-1)^{J_\textup{light}-s}$ (see appendix~\ref{reflap}), making this quantum number redundant. Thus, we label the configurations by the static spin projection $\lambda$ \emph{with sign} and $\eta$, omitting $\epsilon$. We shall denote these static BO quantum numbers that also take into account the spin of the heavy quarks by $\lambda_\eta$, such as $0_g$, $0_u$, $\pm1_g$, $\pm1_u$, $\pm2_g$, $\pm2_u$, and so on.

 In this paper, we shall concentrate on light QCD configurations with BO quantum numbers $\Sigma_g^+$, which can be produced by a $Q\bar{Q}$ source and by di-meson sources. For a $Q\bar{Q}$ source, that is, the product of $Q$ and $\bar{Q}$ fields connected by a gauge transporter, $\Sigma_g^+$ corresponds to $J_\textup{light}=0$, so that $\bm{S}= \bm{S}_{Q\bar{Q}}$. For a di-meson source, that is, a product of $Q\bar{q}$ and $q\bar{Q}$ operators, $\bm{S}$ coincides with the total spin of the di-meson pair. So, for $Q\bar{Q}$ and di-meson sources producing light $\Sigma_g^+$ configurations, $s$ and $\lambda$ are just total spin quantum numbers. Note, however, that for a $Q\bar{Q}$ source  the individual spin quantum numbers are $\frac{1}{2}$, while for a di-meson source the individual spin quantum numbers are integers.

In theory, a $Q\bar{Q}$ source generates all $\Sigma_g^+$ light QCD configurations, labeled by $n=1,2,3,\dots$ in increasing energy levels. However, at any given $Q\bar{Q}$ distance $r=\lvert\bm{r}\rvert$, many of these light QCD configurations will have a very small overlap with the quark-antiquark source, making it difficult to detect them in lattice QCD simulations with finite numerical precision. A practical workaround is to consider the correlation between quark-antiquark and di-meson sources. Following the calculations of the $\Sigma_g^+$ BO potentials with string breaking in lattice QCD \cite{Bal05}, we truncate the operator expansion by considering only di-meson sources for negative-parity mesons, which yields a good description of the ground $n=1$ and the first excited $n=2$ $\Sigma_g^+$ BO configurations.\footnote{In the case of string breaking by light and strange quarks, one can get a good description of the first three configurations by including also di-meson sources for strange negative-parity mesons \cite{Bul19}.}

Here we do  not consider di-meson sources for positive-parity mesons, which are presumably required for the calculation of excited $\Sigma_g^+$ light QCD configurations with larger $n$. Neither do we include configurations with light BO quantum numbers $\Pi_u$, $\Sigma_u^-$, etc., also referred to as hybrid configurations. Hence, the current analysis is restricted to energies well below the lowest threshold for the production of a positive-parity meson ($B\bar{B}_1$) and the energy of the lowest quarkonium hybrid meson.

We shall label the quark-antiquark and di-meson sources producing $\lambda_\eta$ configurations by their total spin $s$ and projection $\lambda$. Note that in the $Q\bar{Q}$, $B\bar{B}$, and $B^\ast\bar{B}^\ast$ cases the quantum number $\eta$ is determined by the total spin of the source as $CP=(-1)^{s+1}$ for quark-antiquark sources and $CP=(-1)^s$ for di-meson sources. On the other hand, one can construct two independent combinations of $B$ and $B^\ast$, $(B\bar{B}^\ast\mp B^\ast\bar{B})/\sqrt{2}$ with $CP=\pm$. In this last case, we shall specify the corresponding quantum number $\eta$ of the meson pair as $B\bar{B}^{\ast\eta}$ whenever necessary. Hence, we shall indicate the $Q\bar{Q}$ sources with $s=0,1$ as $\mathcal{Q}_{s,\lambda}$, the $B\bar{B}$ source with $s=0$ as $\mathcal{B}_{0,0}$, the $B\bar{B}^{\ast\eta}$ sources with $s=1$ as $\mathcal{B}^{\ast\eta}_{1,\lambda}$, and the $B^\ast\bar{B}^\ast$ sources with $s=0,1,2$ as $\mathcal{B}^{\ast\ast}_{s,\lambda}$. We will sometimes use the shorthand $B^{(\ast)}\bar{B}^{(\ast)}$ to indicate all the various meson pairs, $B\bar{B}$, $B\bar{B}^{\ast\eta}$, and $B^\ast\bar{B}^\ast$.

Since $Q$ and $\bar{Q}$ each have two spin states, there are four sets of static BO quantum numbers $\lambda_\eta$ that correspond to the light BO configuration $\Sigma_g^+$. Namely, the four spin states of a $Q\bar{Q}$ pair can be grouped in a triplet corresponding total spin $s=1$, with $\lambda=-1,0,+1$ and $\eta=g$ (that is, $CP=+$), and a singlet corresponding to $s=0$, with $\lambda=0$ and $\eta=u$ (that is, $CP=-$). Therefore, the relevant sets of static BO quantum numbers $\lambda_\eta$ are $0_g$, $+1_g$, $-1_g$, and $0_u$. The corresponding values of the total spin quantum number $s$ for $Q\bar{Q}$ and $B^{(\ast)}\bar{B}^{(\ast)}$ sources are summarized in Table~\ref{quanumtable}. There are also four sets of static BO quantum numbers $\lambda_\eta$ that can be produced by di-meson sources but not quark-antiquark sources: $+2_g$, $-2_g$, $+1_u$, and $-1_u$. Although irrelevant for string breaking, these configurations have also been listed in Table~\ref{quanumtable} as they will become useful later on.

\begin{table}
\centering
\caption{\label{quanumtable}Total spin $s$ for quark-antiquark and di-meson sources producing a light QCD configuration $\Sigma_g^+$ that combines with the spins of the heavy quarks to give the specified static BO quantum numbers $\lambda_\eta$ ($\lambda$ is the total spin projection and $\eta=g,u$ stands for $CP=+,-$).}
\begin{tabular}{cccccc}
\toprule
Source			& $0_g$	& $\pm1_g$	& $\pm2_g$	& $0_u$	& $\pm1_u$	\\
\midrule
$Q\bar{Q}$		& 1		&  1			& n/a		& 0		& n/a 		\\
$B\bar{B}$		& 0		& n/a 		& n/a		& n/a	& n/a 		\\
$B\bar{B}^\ast$		& 1		& 1			& n/a		& 1		& 1			\\
$B^\ast\bar{B}^\ast$	& 0, 2		& 2			& 2			& 1		& 1			\\
\bottomrule
\end{tabular}
\end{table}

We proceed to describe the sources with separation $\bm{r}$ that produce light QCD configurations $\Sigma_g^+$ that when combined with the spins of the heavy quarks have specified static BO quantum numbers $\lambda_\eta$. A quark-antiquark source that creates $0_g$ configurations can be constructed as
\[
\mathcal{Q}_{1,0} = \bar{Q}\bigl(-\tfrac{1}{2}\bm{r}\bigr) \bm{\gamma} \cdot \bm{\hat{r}} \mathcal{W}\bigl(-\tfrac{1}{2}\bm{r}, +\tfrac{1}{2}\bm{r}\bigr) Q\bigl(+\tfrac{1}{2}\bm{r}\bigr)
\]
(see refs.~\cite{Bal05,Bul19}) where $Q(+\frac{1}{2}\bm{r})$ and $\bar{Q}(-\frac{1}{2}\bm{r})$ are the heavy quark and antiquark operators and $\mathcal{W}(-\frac{1}{2}\bm{r}, +\frac{1}{2}\bm{r})$ is the equal-time parallel gauge transporter along the straight line between $-\frac{1}{2}\bm{r}$ and $+\frac{1}{2}\bm{r}$. The $B\bar{B}$ source that creates $0_g$ configurations is
\[
\mathcal{B}_{0,0} = \bigl(\bar{Q}\bigl(-\tfrac{1}{2}\bm{r}\bigr) \gamma_5 q\bigl(-\tfrac{1}{2}\bm{r}\bigl)\bigr) \bigl(\bar{q}\bigl(+\tfrac{1}{2}\bm{r}\bigr)\gamma_5 Q\bigl(+\tfrac{1}{2}\bm{r}\bigr)\bigr)
\]
(see refs.~\cite{Bal05,Bul19}) where $q\bigl(-\frac{1}{2}\bm{r}\bigr)$ and $\bar{q}\bigl(+\frac{1}{2}\bm{r}\bigr)$ are light quark and antiquark operators. Notice that we consider only one flavor of light quarks, for simplicity.

The lattice-QCD calculation of the BO potentials with string breaking in ref.~\cite{Bal05} was carried out using only $Q\bar{Q}$ and $B\bar{B}$ sources in a static BO configuration $0_g$, that is, $\mathcal{Q}_{1,0}$ and $\mathcal{B}_{0,0}$. Although these operators are sufficient to generate the first two $\Sigma_g^+$ energy levels, they do not exhaust all the possible di-meson sources creating a $0_g$ configuration. Moreover, one needs $B\bar{B}^{\ast\eta}$ and $B^\ast\bar{B}^\ast$ sources to create the additional $\pm1_g$ and $0_u$ configurations, which also participate in string breaking; see Table~\ref{quanumtable}. Therefore, we proceed to describe a complete set of quark-antiquark and di-meson sources, including $B\bar{B}^{\ast\eta}$ and $B^\ast\bar{B}^\ast$, which is also instrumental for incorporating the $B^\ast$-$B$ mass splitting later on.

For ease of notation, from here on we shall suppress the argument of the quark operators, in the implicit understanding that $Q$ and $\bar{q}$ are created at $+\frac{1}{2}\bm{r}$, while $\bar{Q}$ and $q$ are created at $-\frac{1}{2}\bm{r}$. We shall also identify the arbitrary axis $\bm{\hat{z}}$ with the $\bm{\hat{r}}$ direction, which is legitimate in the static limit since $\bm{r}$ is a constant. This allows us to simplify expressions involving $\bm{\hat{r}}$ such as $\bm{\gamma} \cdot \bm{\hat{r}} \to\gamma^3$, for instance.

One can construct a gauge-invariant quark-antiquark operator creating light BO quantum numbers $\Sigma_g^+$ by connecting $\bar{Q}$ and $Q$ sources via a straight Wilson line $\mathcal{W}$ and contracting the $\bar{Q}$ and $Q$ spins with a $4\times4$ Dirac matrix. Since the static quark and antiquark sources satisfy $Q = P_+ Q$ and $\bar{Q} = \bar{Q} P_- $ with
\[
P_\pm = \frac{1 \pm \gamma^0}{2},
\]
there are only 4 independent $Q\bar{Q}$ operators in total. There is one spin-0 operator,
\[
\bar{Q} \gamma_5 \mathcal{W} Q,
\]
and three spin-1 operators,
\[
\bar{Q} \gamma^3 \mathcal{W} Q \quad \text{and} \quad \bar{Q} \gamma^\pm \mathcal{W} Q
\]
with spin projection $0$ and $\pm1$, where $\gamma^\pm=\mp(\gamma^1\pm i\gamma^2)/\sqrt{2}$.

On the other hand, one can construct 8 independent operators with the quark content and quantum numbers of a ground-state heavy-light meson. There is the $B$ operator,
\[
\bar{q} \gamma_5 Q,
\]
the three $B^\ast$ operators,
\[
\bar{q} \gamma^3 Q \quad \text{and} \quad \bar{q} \gamma^\pm Q
\]
with spin projection $0$ and $\pm1$, plus the corresponding $\bar{B}$ and $\bar{B}^\ast$ operators that can be obtained by substituting $\bar{q}\to\bar{Q}$ and $Q\to q$. From the product of $B$ and $B^\ast$ operators with $\bar{B}$ and $\bar{B}^\ast$ operators, one can construct 16 $B^{(\ast)}\bar{B}^{(\ast)}$ operators: 1 for $B\bar{B}$, 6 for $B\bar{B}^{\ast \eta}$ (3 for each $\eta=g,u$), and 9 for $B^\ast\bar{B}^\ast$. The 9 $B^\ast\bar{B}^\ast$ operators can be combined in multiplets with definite total spin, that is, a singlet with $s=0$, a triplet with $s=1$, and a quintuplet with $s=2$, using the Clebsch-Gordan coefficients.

These 20 operators, 4 for $Q\bar{Q}$ plus 16 for $B^{(\ast)}\bar{B}^{(\ast)}$, constitute a complete basis for generating static BO configurations associated to the ground and first excited $\Sigma_g^+$ light QCD configurations. Next, we provide explicit expressions for the subset of operators that create $\lambda_\eta$ configurations participating in string breaking. Following Table~\ref{quanumtable}, the $Q\bar{Q}$ and  $B^{(\ast)}\bar{B}^{(\ast)}$ sources that create $0_g$ configurations are
\begin{subequations}
\begin{align}
\mathcal{Q}_{1,0} &= \bar{Q}\gamma^3 \mathcal{W} Q,
\label{qqeq} \\
\mathcal{B}_{0,0} &= \bigl(\bar{Q} \gamma_5 q\bigr) \bigl(\bar{q}\gamma_5 Q\bigr),
\label{bbeq} \\
\mathcal{B}^{\ast g}_{1,0} &= \frac{1}{\sqrt{2}}\bigl[(\bar{Q} \gamma^3 q) (\bar{q} \gamma_5 Q) - (\bar{Q} \gamma_5 q) (\bar{q} \gamma^3 Q)\bigr],
\label{bbasteq} \\
\mathcal{B}^{\ast\ast}_{0,0} &= \frac{1}{\sqrt{3}}\bigl[(\bar{Q} \gamma^- q) (\bar{q} \gamma^+ Q) + (\bar{Q} \gamma^+ q) (\bar{q} \gamma^- Q) - (\bar{Q}\gamma^3 q) (\bar{q} \gamma^3 Q)\bigr],
\label{bbastast0eq} \\
\mathcal{B}^{\ast\ast}_{2,0} &= \frac{1}{\sqrt{6}}\bigl[(\bar{Q} \gamma^- q) (\bar{q} \gamma^+ Q) + (\bar{Q} \gamma^+ q) (\bar{q} \gamma^- Q) + 2 (\bar{Q}\gamma^3 q) (\bar{q} \gamma^3 Q) \bigr],
\label{bbastast2eq}
\end{align}
\end{subequations}
The $Q\bar{Q}$ and $B^{(\ast)}\bar{B}^{(\ast)}$ sources that create $\pm1_g$ configurations are
\begin{subequations}
\begin{align}
\mathcal{Q}_{1,\pm1} &= \bar{Q}\gamma^\pm \mathcal{W} Q,
\label{qqeq2}\\
\mathcal{B}^{\ast g}_{1,\pm1} &= \frac{1}{\sqrt{2}}\bigl[(\bar{Q} \gamma^\pm q) (\bar{q} \gamma_5 Q) - (\bar{Q} \gamma_5 q) (\bar{q} \gamma^\pm Q)\bigr],
\label{bbasteq2}\\
\mathcal{B}^{\ast\ast}_{2,\pm1} &= \frac{1}{\sqrt{2}}\bigl[(\bar{Q} \gamma^3 q) (\bar{q} \gamma^\pm Q) + (\bar{Q} \gamma^\pm q) (\bar{q} \gamma^3 Q)\bigr].
\label{bbastasteq2}
\end{align}
\end{subequations}
The $Q\bar{Q}$ and $B^{(\ast)}\bar{B}^{(\ast)}$ sources that create $0_u$ configurations are
\begin{subequations}
\begin{align}
\mathcal{Q}_{0, 0} &= \bar{Q}\gamma_5 \mathcal{W} Q,
\label{qqeq3} \\
\mathcal{B}^{\ast u}_{1, 0} &= \frac{1}{\sqrt{2}}\bigl[(\bar{Q} \gamma^3 q) (\bar{q} \gamma_5 Q) + (\bar{Q} \gamma_5 q) (\bar{q} \gamma^3 Q)\bigr],
\label{bbasteq3} \\
\mathcal{B}^{\ast\ast}_{1,0} &= \frac{1}{\sqrt{2}}\bigl[(\bar{Q} \gamma^- q) (\bar{q} \gamma^+ Q) - (\bar{Q} \gamma^+ q) (\bar{q} \gamma^- Q)\bigr].
\label{bbastasteq3}
\end{align}
\end{subequations}

\section{String breaking and transition rates}
\label{stringsec}

In lattice QCD, string breaking is manifested explicitly as a nonvanishing transition rate between states created by quark-antiquark and di-meson sources. Lattice QCD has been used to calculate the transition rate between $\Sigma_g^+$ states created by $Q\bar{Q}$ and $B\bar{B}$ sources in a $0_g$ static BO configuration \cite{Bal05}. In this section, we express the transition rates between $Q\bar{Q}$ and $B^{(\ast)}\bar{B}^{(\ast)}$ sources for different $\lambda_\eta$ in terms of the string-breaking transition rate $g$ between a gluonic string with light BO quantum numbers $\Sigma_g^+$ and a light quark-antiquark pair within the same configuration.

Any di-meson operator can be factorized into products of light-quark operators and heavy-quark operators using the Fierz identity,
\[
q \bar{q} = -\frac{1}{4}\biggl[ (\bar{q} q) + (\bar{q} \gamma^\mu q) \gamma_\mu + \frac{1}{2} (\bar{q} \sigma^{\mu\nu} q) \sigma_{\mu\nu} - (\bar{q} \gamma^\mu \gamma_5 q) \gamma_\mu \gamma_5 + (\bar{q} \gamma_5 q) \gamma_5 \biggr]
\]
where the overall minus sign comes from the anticommutation of fermionic operators. Specifically, the di-meson sources in eqs.~\eqref{bbeq}-\eqref{bbastast2eq} that create static BO configurations $0_g$ become
\begin{subequations}
\begin{align}
\mathcal{B}_{0,0} =& \frac{1}{2} \bigl[(\bar{q} P_- \gamma^+ q) (\bar{Q} \gamma^- Q) + (\bar{q} P_- \gamma^- q) (\bar{Q} \gamma^+ Q) - (\bar{q} P_- \gamma^3 q) (\bar{Q} \gamma^3 Q) \nonumber \\
&\hphantom{\frac{1}{2} \bigl[}- (\bar{q} P_- \gamma_5 q) (\bar{Q} \gamma_5 Q)\bigr], \label{beq}\\
\mathcal{B}^{\ast g}_{1,0} =& \frac{1}{\sqrt{2}}\bigl[(\bar{q} P_- \gamma^+ q) (\bar{Q} \gamma^- Q) - (\bar{q} P_- \gamma^- q) (\bar{Q} \gamma^+ Q)\bigr], \label{b*eq}\\
\mathcal{B}^{\ast\ast}_{0,0} =& \frac{1}{2\sqrt{3}}\bigl[(\bar{q} P_- \gamma^+ q) (\bar{Q} \gamma^- Q) + (\bar{q} P_- \gamma^- q) (\bar{Q} \gamma^+ Q) - (\bar{q} P_- \gamma^3 q) (\bar{Q} \gamma^3 Q)\nonumber\\
&\hphantom{\frac{1}{2\sqrt{3}}\bigl[} + 3 (\bar{q} P_- \gamma_5 q) (\bar{Q} \gamma_5 Q)\bigr], \label{b**0eq}\\
\mathcal{B}^{\ast\ast}_{2,0} =& -\frac{1}{\sqrt{6}}\bigl[(\bar{q} P_- \gamma^+ q) (\bar{Q} \gamma^- Q) + (\bar{q} P_- \gamma^- q) (\bar{Q} \gamma^+ Q) + 2 (\bar{q} P_- \gamma^3 q) (\bar{Q} \gamma^3 Q)\bigr]. \label{b**2eq}
\end{align}
\label{beqs}
\end{subequations}

HQSS implies that the $Q$ and $\bar{Q}$ spins are conserved. The 4 operators $\bar{Q} \gamma^3 Q$, $\bar{Q} \gamma^+ Q$, $\bar{Q} \gamma^- Q$, and  $\bar{Q} \gamma_5 Q$ in eqs.~\eqref{beqs} correspond to 4 independent $Q\bar{Q}$ spin states. In the quark-antiquark source $\mathcal{Q}_{1,0}$ in eq.~\eqref{qqeq}, the $Q\bar{Q}$ spin state is given by the operator $\bar{Q}\gamma^3 Q$. Hence, the static correlation function between $\mathcal{Q}_{1,0}$ and any of the di-meson sources in eqs.~\eqref{beqs} proceeds exclusively through the terms in the latter that are proportional to $\bar{Q} \gamma^3 Q$,
\begin{equation}
\braket{0| \mathcal{B}_{s,0}^{(\ast\ast)} \mathcal{T}_\tau \mathcal{Q}_{1,0}^\dag | 0} \propto \braket{0| (\bar{q} P_- \gamma^3 q) (\bar{Q} \gamma^3 Q) \mathcal{T}_\tau (\bar{Q}\gamma^3 \mathcal{W} Q)^\dag | 0}
\label{correq}
\end{equation}
where $\mathcal{B}_{s,0}^{(\ast\ast)}$ is a shorthand for any of the di-meson sources in eqs.~\eqref{beqs} and $\mathcal{T}_\tau$ is the Euclidean time-evolution operator from time $0$ to $\tau$. Factoring out the $Q\bar{Q}$ spin given by $\bar{Q}\gamma^3Q$, the correlator on the right side of eq.~\eqref{correq} can be cast symbolically as $\braket{0 |(\bar{q} P_- \gamma^3 q) \mathcal{T}_t \mathcal{W}^\dag | 0}$. From it, the string-breaking transition rate $g$ can be obtained by evaluating the time derivative with an appropriate normalization coefficient after some relaxation time for which higher excited light QCD configurations have decayed; see ref.~\cite{Bal05}.

Thus, it follows from eq.~\eqref{correq} that the transition rates between $Q\bar{Q}$ and $B^{(\ast)}\bar{B}^{(\ast)}$ are all proportional to the string-breaking transition rate $g$. For the $0_g$ configuration, the coefficients are simply the numerical factors multiplying $\bar{Q} \gamma^3 Q$ inside eqs.~\eqref{beqs}. The transition rates then read
\begin{subequations}
\begin{align}
G_{Q\bar{Q}(1),B\bar{B}(0)}^{g,0}(r) &= -\frac{1}{2}g(r),\\
G_{Q\bar{Q}(1),B\bar{B}^\ast(1)}^{g,0}(r) &= 0,\\
G_{Q\bar{Q}(1),B^\ast\bar{B}^\ast(0)}^{g,0}(r) &= -\frac{1}{2\sqrt{3}}g(r),\\
G_{Q\bar{Q}(1),B^\ast\bar{B}^\ast(2)}^{g,0}(r) &= -\sqrt{\frac{2}{3}} g(r),
\end{align}
\label{sg+eqs}
\end{subequations}
where we have specified the static BO quantum numbers $\eta$ and $\lambda$ of the configuration as superscripts, the total spin of the quark-antiquark and di-meson sources as $Q\bar{Q}(s)$ and $B^{(\ast)}\bar{B}^{(\ast)}(s)$, respectively, and the distance $r$ between the sources as an argument.

One can proceed in the exact same manner for the $\pm1_g$ and $0_u$ configurations, by considering the corresponding quark-antiquark and di-meson sources. For the $\pm1_g$ configurations, the di-meson sources in eqs.~\eqref{bbasteq2}-\eqref{bbastasteq2} become
\begin{subequations}
\begin{align}
\mathcal{B}^{\ast g}_{1,\pm1} &= \pm\frac{1}{\sqrt{2}}\bigl[(\bar{q} P_- \gamma^\pm q) (\bar{Q} \gamma^3 Q) - (\bar{q} P_- \gamma^3 q) (\bar{Q} \gamma^\pm Q)\bigr], \\
\mathcal{B}^{\ast\ast}_{2,\pm1} &= -\frac{1}{\sqrt{2}}  \bigl[(\bar{q} P_- \gamma^\pm q) (\bar{Q} \gamma^3 Q) + (\bar{q} P_- \gamma^3 q) (\bar{Q} \gamma^\pm Q)\bigr].
\end{align}
\label{beqs2}
\end{subequations}
Since the spin states of the heavy quark-antiquark pair in $\mathcal{Q}_{1,\pm1}$ in eq.~\eqref{qqeq2} are given by $\bar{Q} \gamma^\pm Q$, the transition rates are
\begin{subequations}
\begin{align}
G_{Q\bar{Q}(1),B\bar{B}^\ast(1)}^{g,\pm1}(r) &= \mp \frac{1}{\sqrt{2}}g(r),\\
G_{Q\bar{Q}(1),B^\ast\bar{B}^\ast(2)}^{g,\pm1}(r) &= -\frac{1}{\sqrt{2}}g(r).
\end{align}
\label{pgeqs}
\end{subequations}
For the $0_u$ configuration, the di-meson sources in eqs.~\eqref{bbasteq3}-\eqref{bbastasteq3} become
\begin{subequations}
\begin{align}
\mathcal{B}^{\ast u}_{1, 0} &= -\frac{1}{\sqrt{2}}\bigl[(\bar{q} P_- \gamma_5 q) (\bar{Q} \gamma^3 Q) + (\bar{q} P_- \gamma^3 q) (\bar{Q} \gamma_5 Q)\bigr], \\
\mathcal{B}^{\ast\ast}_{1,0} &= \frac{1}{\sqrt{2}}  \bigl[(\bar{q} P_- \gamma_5 q) (\bar{Q} \gamma^3 Q) - (\bar{q} P_- \gamma^3 q) (\bar{Q} \gamma_5 Q)\bigr].
\end{align}
\label{beqs3}
\end{subequations}
Since the spin state of the heavy quark-antiquark pair in $\mathcal{Q}_{0,0}$ in eq.~\eqref{qqeq3} is given by $\bar{Q} \gamma_5 Q$, the transition rates are
\begin{subequations}
\begin{align}
G_{Q\bar{Q}(0),B\bar{B}^\ast(1)}^{u,0}(r) &= -\frac{1}{\sqrt{2}}g(r),\\
G_{Q\bar{Q}(0),B^\ast\bar{B}^\ast(1)}^{u,0}(r) &= -\frac{1}{\sqrt{2}}g(r).
\end{align}
\label{su-eqs}
\end{subequations}

As explained in more detail in the next section, one can use eqs.~\eqref{sg+eqs}-\eqref{su-eqs} to verify that, for all four $\lambda_\eta$ configurations, a unitary change of basis reduces the interaction matrix between $Q\bar{Q}$ and $B^{(\ast)}\bar{B}^{(\ast)}$ down to a $2\times 2$ string-breaking interaction sub-matrix $\bm{G}(r)$ between $\mathcal{W}$ and $\bar{q} P_- \gamma^3 q$, plus rows and columns associated to decoupled $\bar{q} P_- \gamma^\pm q$ and $\bar{q} P_- \gamma_5 q$ components. Otherwise said, the relevant string-breaking interaction matrix is the same independently of the spins of the heavy quarks, consistently with HQSS.

The $2\times2$ string-breaking interaction matrix reads
\begin{equation}
\bm{G}(r) = 
\begin{pmatrix}
V_ {\mathcal{Q}}(r)	& g(r)	\\
g(r)		& 0	\\
\end{pmatrix},
\label{cormateq}
\end{equation} 
where $V_ {\mathcal{Q}}(r)$ is the $Q\bar{Q}$ static energy relative to twice the static $B$ meson mass. Notice that the zero in the lower right corner of eq.~\eqref{cormateq} corresponds to approximating the $B^{(\ast)}\bar{B}^{(\ast)}$ static energy by its constant value at large $r$. The constant can be set to zero by taking the zero of the energy to be the static meson pair threshold. This ansatz provides a good fit to the lattice QCD data in \cite{Bal05}; see, for instance, ref.~\cite{Bic20}.

It is worth mentioning that the results presented here are straightforwardly extended to the case of dynamical quark fields with $N_f$ flavors and $SU(N_f)$ flavor symmetry. The only notable difference is that the string-breaking transition rate with $N_f$ light flavors is equal to the single-flavor one times a factor $\sqrt{N_f}$. If, alternatively, both light and strange quarks are considered \cite{Bul19}, then one needs to substitute eq.~\eqref{cormateq} with an appropriate $3\times 3$ string-breaking interaction matrix, like
\begin{equation}
\bm{G}(r) =
\begin{pmatrix}
V_ {\mathcal{Q}}(r)	& g(r)	& g_s(r)		\\
g(r)			& 0		& 0			\\
g_s(r)		& 0		& 2 \delta_s	\\
\end{pmatrix}
\label{modeleq}
\end{equation}
where $g_s(r)$ is the string-breaking transition rate by strange quarks and $\delta_s = m_{B_s} -  m_{B}$ is the mass difference between static $B_s$ and $B$ mesons. Notice that we have assumed that the $B^{(\ast)}\bar{B}^{(\ast)}$ and $B_s^{(\ast)}\bar{B}_s^{(\ast)}$ static energies are well-approximated by their constant values at large $r$. We have also assumed that the transition rate between $B^{(\ast)}\bar{B}^{(\ast)}$ and $B_s^{(\ast)}\bar{B}_s^{(\ast)}$ sources is negligible, since it requires the creation of $s\bar{s}$ which is OZI suppressed. The ansatz in eq.~\eqref{modeleq} with $g(r)$ and $g_s(r)$ replaced by constants is identical to the parametrization used in ref.~\cite{Bul19} for fitting the lattice QCD interaction matrix with $2+1$ light-quark flavors.

\section{Threshold spin splittings}
\label{threshsec}

To go beyond the static approximation, one has to introduce corrections in increasing powers of $1/m_{Q}$ that typically depend on the heavy-quark momenta and/or spins. We shall limit ourselves to the first-order corrections, that is, the nonrelativistic kinetic energies of the heavy quarks and heavy-quark spin effects of first order in $1/m_Q$. In this section we shall focus on the latter, leaving the treatment of the kinetic energy for later.

At first order in $1/m_Q$, the heavy-quark spin corrections to the static approximation consist of couplings between the angular momentum of the light QCD fields and the spins of the heavy quarks \cite{Bra19b}. For the $Q\bar{Q}$ configurations with light BO quantum numbers $\Sigma_g^+$ ($J_\textup{light}=0$), these $1/m_Q$ couplings vanish. For the $B^{(\ast)}\bar{B}^{(\ast)}$ configurations, one can apply heavy-quark effective theory to calculate these $1/m_Q$ effects for each individual heavy-light meson, which are responsible for the $B^\ast$-$B$ mass splitting \cite{Wis92,Fal93,Boy95,Bra18b}. As for the $Q\bar{Q}$-$B^{(\ast)}\bar{B}^{(\ast)}$ transition rates, we shall assume that they are reasonably approximated by their HQSS expressions in eqs.~\eqref{sg+eqs}, \eqref{pgeqs}, and \eqref{su-eqs} at first order in $1/m_Q$. Therefore, for the current study, taking into account heavy-quark spin corrections of first order in $1/m_Q$ boils down to splitting the $B\bar{B}$, $B\bar{B}^{\ast\eta}$, and $B^\ast\bar{B}^\ast$ threshold masses.

Threshold spin splittings are easily included in the interaction matrix between $Q\bar{Q}$ and $B^{(\ast)}\bar{B}^{(\ast)}$ in each $\lambda_\eta$ configuration. Let $\Delta=m_{B^\ast} - m_B$ be the mass difference between $B^\ast$ and $B$ mesons and $m_{B^{(\ast)}}=(m_B + 3 m_{B^\ast})/4$ the spin-average mass. Taking, for instance, the $0_g$ configuration, the corrected interaction matrix between $\mathcal{Q}_{1,0}$, $\mathcal{B}_{0,0}$, $\mathcal{B}^{\ast g}_{1,0}$, $\mathcal{B}^{\ast\ast}_{0,0}$, and $\mathcal{B}^{\ast\ast}_{2,0}$ is the $5\times5$ matrix
\begin{equation}
\bm{G}^{g,0}(r) = 
\begin{pmatrix}
V_ {\mathcal{Q}}(r)		& \bm{v} \, g(r)	\\
\bm{v}^\dag \, g(r)	& \bm{\mathfrak{M}}	\\
\end{pmatrix}
\label{g0mesmateq}
\end{equation}
where $\bm{v}$ is the unit row vector of coefficients from eqs.~\eqref{sg+eqs},
\[
\bm{v} = 
\begin{pmatrix}
-\frac{1}{2}	& 0		& -\frac{1}{2\sqrt{3}}	& -\sqrt{\frac{2}{3}}	\\
\end{pmatrix},
\]
and
\[
\bm{\mathfrak{M}} = \diag(-3, -1, 1, 1) \frac{\Delta}{2},
\]
is the $4\times4$ threshold mass matrix relative to the spin-average $B^{(\ast)}\bar{B}^{(\ast)}$ threshold,
\[
2 m_{B^{(\ast)}} = \frac{m_B + 3 m_{B^\ast}}{2}.
\]

Notice that, in the HQSS limit $\Delta\to0$, the $5\times5$ matrix $\bm{G}^{g,0}(r)$ in eq.~\eqref{g0mesmateq} can be reduced to the $2\times2$ matrix $\bm{G}(r)$ in eq.~\eqref{cormateq} plus vanishing rows and columns by a unitary change of basis for the 4 di-meson channels that transforms $\bm{v}$ into $(1,0,0,0)$. However, when $\Delta>0$ the threshold mass matrix $\bm{\mathfrak{M}}$ is also affected by this transformation, such that one cannot have both a diagonal $\bm{\mathfrak{M}}$ and $\bm{v}=(1,0,0,0)$ at the same time. This shows that string breaking with threshold spin splittings breaks HQSS.

This HQSS-breaking effect can also be pointed out by calculating the interaction matrices in the $\pm1_g$ and $0_u$ configurations. For the $\pm1_g$ configurations, the interaction matrices between $\mathcal{Q}_{1,\pm1}$, $\mathcal{B}^{\ast g}_{1,\pm1}$, and $\mathcal{B}^{\ast\ast}_{2,\pm1}$ are $3\times3$ matrices with the off-diagonal entries from eqs.~\eqref{pgeqs},
\begin{equation}
\bm{G}^{g,\pm1}(r) =
\begin{pmatrix}
V_ {\mathcal{Q}}(r)		& \mp\frac{1}{\sqrt{2}}g(r)	& -\frac{1}{\sqrt{2}}g(r)	\\
\mp\frac{1}{\sqrt{2}}g(r)	& -\frac{1}{2}\Delta			& 0				\\
-\frac{1}{\sqrt{2}}g(r)	& 0						& \frac{1}{2}\Delta			\\
\end{pmatrix}.
\label{gpm1mesmateq}
\end{equation}
For the $0_u$ configuration, the interaction matrix between $\mathcal{Q}_{0,0}$, $\mathcal{B}^{\ast u}_{1,0}$, and $\mathcal{B}^{\ast\ast}_{1,0}$ is a $3\times3$ matrix with the off-diagonal entries from eqs.~\eqref{su-eqs},
\begin{equation}
\bm{G}^{u,0}(r) =
\begin{pmatrix}
V_ {\mathcal{Q}}(r)		& -\frac{1}{\sqrt{2}}g(r)	& -\frac{1}{\sqrt{2}}g(r)	\\
-\frac{1}{\sqrt{2}}g(r)	& -\frac{1}{2}\Delta				& 0			\\
-\frac{1}{\sqrt{2}}g(r)	& 0						& \frac{1}{2}\Delta		\\
\end{pmatrix}.
\label{u0mesmateq}
\end{equation}

Notice that, in the HQSS limit $\Delta\to0$, the $3\times3$ interaction matrices $\bm{G}^{g,\pm1}(r)$ and $\bm{G}^{u,0}(r)$ in eqs.~\eqref{gpm1mesmateq} and \eqref{u0mesmateq} can be reduced to the same $2\times2$ matrix $\bm{G}(r)$ in eq.~\eqref{cormateq} plus vanishing rows and columns by a unitary change of basis for the 2 di-meson channels.

Note that $B\bar{B}^{\ast u}$ and $B^\ast\bar{B}^\ast$ can also produce $\lambda_\eta$ configurations that have no overlap with those created by $Q\bar{Q}$; see Table~\ref{quanumtable}. Such configurations yield trivial ``interaction matrices'' which are just the threshold mass matrices for the corresponding di-meson channels. As will be shown later, a formal BO approximation must include the interaction matrices for all the $\lambda_\eta$ configurations that can be produced by the sources involved. Hence, for the sake of completeness, we list the interaction matrices for the $\pm2_g$ and $\pm1_u$ configurations as well. For $\pm2_g$, one has the $1\times1$ matrices
\begin{equation}
\bm{G}^{g,\pm2}(r) =
\begin{pmatrix}
\frac{1}{2}\Delta
\end{pmatrix}
\label{gpm2mesmateq}
\end{equation}
involving only $\mathcal{B}^{\ast\ast}_{2,\pm2}$. For $\pm1_u$, one has the diagonal $2\times2$ matrices
\begin{equation}
\bm{G}^{u,\pm1}(r) = \diag(-1, 1) \frac{\Delta}{2}
\label{upm1mesmateq}
\end{equation}
between $\mathcal{B}^{\ast u}_{1,\pm1}$ and $\mathcal{B}^{\ast\ast}_{1,\pm1}$.

\section{Diabatic Born-Oppenheimer approximation}
\label{bosec}

\subsection{Introducing the kinetic energy}

The other first-order correction in $1/m_Q$ to the static approximation is the nonrelativistic kinetic energy for the relative motion of the heavy quark-antiquark pair,
\begin{equation}
K= p^2 / m_Q.
\label{kineq}
\end{equation}
The introduction of the kinetic energy operator gives motion to the heavy sources, which can then be determined by solving a coupled-channel Schr\"odinger equation. This is the core idea of the BO approximation. In this section, we shall detail the formal derivation of this master Schr\"odinger equation in the the \emph{diabatic representation} of BO, which allows one to study heavy quark-antiquark systems using the interaction matrix from \textit{ab initio} lattice QCD.

As we shall see next, one cannot naively plug the lattice-QCD interaction matrix in eq.~\eqref{cormateq} or one of its corrected forms in eqs.~\eqref{g0mesmateq}-\eqref{u0mesmateq} as a potential inside a coupled-channel Schr\"odinger equation. Instead, what governs the motion of the heavy quarks is the \emph{diabatic potential matrix}, which is constructed from a combination of the interaction matrices corresponding to different $\lambda_\eta$ configurations.

Formally, the interaction matrix from lattice QCD can be embedded in a quantum-mechanical framework using the following procedure. Let us indicate the interaction matrix with quark-antiquark distance $r$ within the static BO configuration $\lambda_\eta$ as $\bm{G}^{\eta,\lambda}(r)$. Let $V_n^{\eta,\lambda}(r)$ be the $n$th eigenvalue of $\bm{G}^{\eta,\lambda}(r)$, that is, the $n$th static energy level with static BO quantum numbers $\lambda_\eta$, and $\ket{\zeta_n^{\eta,\lambda}(\bm{r})}$ the corresponding static eigenstate. One can then construct a first-quantized operator,
\begin{equation}
H_\textup{static}(\bm{r}) = \sum_{\eta,\lambda,n} \ket{\zeta_n^{\eta,\lambda}(\bm{r})} V_n^{\eta,\lambda}(r) \bra{\zeta_n^{\eta,\lambda}(\bm{r})},
\label{lighteq}
\end{equation}
acting as a Hamiltonian for the light QCD fields in presence of a heavy quark-antiquark pair at relative position $\bm{r}$. From it, the BO Hamiltonian is constructed as
\[
H_\textup{BO} = K + H_\textup{static},
\]
so that a quantum state $\ket{\Psi}$ of the heavy quark-antiquark system corresponds to a solution of the eigenvalue equation
\begin{equation}
H_\textup{BO}\ket{\Psi} = E\ket{\Psi}
\label{evaleq}
\end{equation}
with $E$ the energy relative to the spin-average $B^{(\ast)}\bar{B}^{(\ast)}$ threshold.

Notice that, whilst the static energy levels $V_n^{\eta,\lambda}(r)$ depend on the distance $r=\lvert \bm{r} \rvert$ alone, the static eigenstates $\ket{\zeta_n^{\eta,\lambda}(\bm{r})}$ depend on the direction $\bm{\hat{r}} = \bm{r} / r$ as well. Most importantly, $\lambda$ itself depends on $\bm{\hat{r}}$ through its definition $\lambda=\bm{S}\cdot\bm{\hat{r}}$. In this case one cannot simply identify the fixed axis $\bm{\hat{z}}$ with $\bm{\hat{r}}$, since $\bm{r}$ is a dynamical variable.

Translating eq.~\eqref{evaleq} into a differential equation in $\bm{r}$ requires an appropriate expansion of the heavy quark-antiquark state $\ket{\Psi}$ in terms of some basis for the light QCD fields. One natural choice is the basis formed by the static eigenstates $\ket{\zeta_n^{\eta,\lambda}(\bm{r})}$, called \emph{adiabatic basis}, since it diagonalizes the light QCD energy operator in eq.~\eqref{lighteq}. Another convenient choice is a basis that diagonalizes the kinetic energy operator in eq.~\eqref{kineq}, called \emph{diabatic basis} \cite{Lic63,Smi69}.

In general, it is possible to express each element of some diabatic basis as a linear combination of the static eigenstates,
\begin{equation}
\ket{\zeta_i^{\eta,\sigma}(\bm{r}, \bm{r}_0)} = \sum_{\lambda,n} A_{n,i}^{\lambda,\sigma}(\bm{r},\bm{r}_0) \ket{\zeta_n^{\eta,\lambda}(\bm{r})}
\label{diabdefeq}
\end{equation}
where the coefficients $A_{n,i}^{\lambda,\sigma}(\bm{r},\bm{r}_0)$ form a unitary matrix $\bm{A}(\bm{r},\bm{r}_0)$ which can be obtained by solving a first-order differential equation with boundary conditions at $\bm{r}=\bm{r}_0$; see, for instance, refs.~\cite{Bru20,Bae06}. A conventional choice for the boundary conditions is to match the diabatic basis to the adiabatic one at $\bm{r}=\bm{r}_0$,
\[
\ket{\zeta_i^{\eta,\sigma}(\bm{r}=\bm{r}_0, \bm{r}_0)} = \sum_{\lambda,n} \delta_{\lambda,\sigma} \delta_{n, i}  \ket{\zeta_n^{\eta,\lambda}(\bm{r}=\bm{r}_0)},
\]
which is the same as imposing 
\[
\bm{A}(\bm{r}=\bm{r}_0,\bm{r}_0) = \bm{I}
\]
with $\bm{I}$ for the identity matrix. It is important to realize that for $\bm{\hat{r}}=\bm{\hat{r}}_0$ the cylindrical symmetry of BO implies
\begin{equation}
A_{n,i}^{\lambda,\sigma}(\bm{r}=r \bm{\hat{r}}_0,\bm{r}_0) = \delta_{\lambda,\sigma}A_{n,i}(r, r_0),
\label{simpleaeq}
\end{equation}
where $r_0=\lvert\bm{r}_0\lvert$ and $A_{n,i}(r, r_0)$ are the coefficients of a radial unitary matrix $\bm{A}(r, r_0)$ which can be obtained by solving a first-order differential equation in $r$ with boundary conditions $\bm{A}(r=r_0, r_0) = \bm{I}$. Notice that the right side of eq.~\eqref{simpleaeq} does not depend on $\bm{\hat{r}}_0$.

Different choices of the matching point $\bm{r}_0$ with the adiabatic basis yield different but equivalent choices of diabatic basis. For the present study, it is most convenient to fix $\bm{r}_0=0$. The reason is that the diabatic basis in this case is matched to the light QCD eigenstates calculated in the static limit where the two sources are placed at the same point $\bm{r}=0$. Unlike the general case, where a nonvanishing separation vector $\bm{r}$ sets a preferred direction in space, the static limit with $\bm{r}=0$ is symmetric under $O(3)\otimes C$. So, the static eigenstates at $\bm{r}=0$ can be classified by the static spin quantum number $s$ and by the $J^{PC}_\textup{light}$ quantum numbers of the light QCD fields, in addition to the static BO quantum numbers $\lambda_\eta$.

As mentioned in section~\ref{statsec}, we restrict our analysis to the ground and first-excited $\Sigma_g^+$ light BO configurations generated by a $Q\bar{Q}$ source and by di-meson sources for negative-parity mesons. The $Q\bar{Q}$ operators for $\bm{r}=0$  generate $J^{PC}_\textup{light}=0^{++}_\textup{light}$; see eqs.~\eqref{qqeq}, \eqref{qqeq2}, and \eqref{qqeq3}. On the other hand, the $B^{(\ast)}\bar{B}^{(\ast)}$ operators for $\bm{r}=0$ generate $J^{PC}_\textup{light}=1^{--}_\textup{light}$ and $0^{-+}_\textup{light}$; see eqs.~\eqref{beqs}, \eqref{beqs2}, and \eqref{beqs3}. The weaker cylindrical symmetry for $\bm{r}\ne0$ allows the $0^{++}_\textup{light}$ component of the $Q\bar{Q}$ configurations to mix with the $1^{--}_\textup{light}$ ($\Lambda=0$) component of the $B^{(\ast)}\bar{B}^{(\ast)}$ configurations through string breaking; see section~\ref{stringsec}. However, such mixing is forbidden by the stronger $O(3)\otimes C$ symmetry for $\bm{r}=0$.\footnote{From this argument, it follows that the string-breaking transition rate $g(r)$ must go to zero as $r\to0$.} Paradoxically, the absence of mixing for $\bm{r}=0$ is what makes $\bm{r}_0=0$ the ideal choice of diabatic basis for studying string breaking. In fact, this choice ensures that the mixing problem is clearly organized in terms of channels with well-defined particle number and spin. Hence, we work with a truncated diabatic basis which consists of 4 $Q \bar Q$ and 16 $B^{(\ast)} \bar B^{(\ast)}$ channels labeled by their total spin $s$ and the diabatic BO quantum numbers $\sigma_\eta$.

This truncated diabatic basis can be formally written as a set $\{\ket{\zeta_i^{\eta,\sigma}(\bm{r},0)}\}$ labeled by $\sigma_\eta$ and the additional label $i$ that specifies the quark-antiquark or di-meson channels with total spin $s$ as $Q\bar{Q}(s)$ or $B^{(\ast)}\bar{B}^{(\ast)}(s)$. Following Table~\ref{quanumtable}, there are five channels in the $0_g$ configuration,
\begin{equation}
0_g \colon i \in \Set{Q\bar{Q}(1), B\bar{B}(0),B\bar{B}^\ast(1),B^\ast\bar{B}^\ast(0),B^\ast\bar{B}^\ast(2)},
\label{diabsiggeq}
\end{equation}
three channels in each of the $+1_g$ and $-1_g$ configurations,
\begin{equation}
\pm1_g \colon i \in \Set{Q\bar{Q}(1),B\bar{B}^\ast(1),B^\ast\bar{B}^\ast(2)},
\label{diabpigeq}
\end{equation}
and three channels in the $0_u$ configuration,
\begin{equation}
0_u \colon i \in \Set{Q\bar{Q}(0), B\bar{B}^\ast(1),B^\ast\bar{B}^\ast(1)}.
\label{diabsigueq}
\end{equation}
On top of these, one should take into account the channels in $\sigma_\eta$ configurations that can be produced by di-meson sources only. Each of the $+2_g$ and $-2_g$ configurations has only one channel,
\begin{equation}
\pm2_g \colon i\in\set{B^\ast\bar{B}^\ast(2)},
\label{diabdelgeq}
\end{equation}
and each of the $+1_u$ and $-1_u$ configurations has two channels,
\begin{equation}
\pm1_u \colon i\in\set{B\bar{B}^\ast(1),B^\ast\bar{B}^\ast(1)}.
\label{diabpiueq}
\end{equation}
Notice that the superscript $\eta$ in $B\bar{B}^{\ast\eta}$ is now redundant, hence we have suppressed it.

The expansion of the heavy quark-antiquark state $\ket{\Psi}$ in this diabatic basis reads
\begin{equation}
\ket{\Psi} = \sum_{\eta,i,\sigma} \int \mathrm{d}^3\bm{r} \, \Psi_i^{\eta,\sigma}(\bm{r}) \ket{\bm{r}} \ket{\zeta_i^{\eta,\sigma}(\bm{r}, 0)},
\label{diabexpeq}
\end{equation}
with the expansion coefficients $\Psi_i^{\eta,\sigma}(\bm{r})$ acting as wave function components for the various diabatic channels. The eigenvalue equation for the BO Hamiltonian, eq.~\eqref{evaleq}, translates into a coupled-channel Schr\"odinger equation,
\begin{equation}
\sum_{i^\prime,\sigma^\prime}\biggl(-\delta_{i,i^\prime}\delta_{\sigma,\sigma^\prime}\frac{\nabla^2}{m_Q} + V_{i,i^\prime}^{\eta,\sigma,\sigma^\prime}(\bm{r})\biggr) \Psi_{i^\prime}^{\eta,\sigma^\prime}(\bm{r}) = E \Psi_i^{\eta,\sigma}(\bm{r})
\label{schreq}
\end{equation}
where $\eta$ is conserved and
\begin{equation}
V_{i,i^\prime}^{\eta,\sigma,\sigma^\prime}(\bm{r}) = \braket{\zeta_i^{\eta,\sigma}(\bm{r},0) | H_\textup{static}(\bm{r}) | \zeta_{i^\prime}^{\eta,\sigma^\prime}(\bm{r},0)}
\label{potdefeq}
\end{equation}
are the elements of the diabatic potential matrix $\bm{V}^\eta(\bm{r})$. The dimension of the matrix $\bm{V}^\eta(\bm{r})$ can be determined through eqs.~\eqref{diabsiggeq}-\eqref{diabpiueq} by counting the total number of channels from all the $\sigma_\eta$ configurations sharing the same quantum number $\eta$. Thus, the diabatic potential matrix $\bm{V}^g(\bm{r})$ within $CP=+$ is a $13\times13$ matrix and the diabatic potential matrix $\bm{V}^u(\bm{r})$ within $CP=-$ is a $7\times7$ matrix.

\subsection{Diabatic potential matrix}

Next, we show that the diabatic potential matrix $\bm{V}^\eta(\bm{r})$ is completely determined by a combination of the interaction matrices $\bm{G}^{\eta,\lambda}(r)$ with various $\lambda$. Let us begin by observing that a spin state with fixed projection onto a general direction $\bm{\hat{r}}$ can be expressed as a Wigner rotation of the corresponding spin state with fixed projection along $\bm{\hat{z}}$ \cite{Wig59}. Thus, one can expand the spin state $\ket{s, \bm{S}\cdot\bm{\hat{r}}=\lambda}$ in terms of the canonical spin states $\ket{s,\sigma}$ as
\begin{equation}
\ket{s, \bm{S}\cdot\bm{\hat{r}}=\lambda} = \sum_\sigma D_{\sigma,\lambda}^s(\varphi,\theta,\psi) \ket{s,\sigma}
\label{roteq}
\end{equation}
with $D_{\sigma,\lambda}^s$ the Wigner $D$-matrix element and $(\varphi,\theta,\psi)$ the Euler angles, where $\theta$ and $\varphi$ are identified with the polar angles of $\bm{\hat{r}}$ while $\psi$ is left arbitrary. Since the static eigenstates at $\bm{r}=0$ have definite spin $s$ and projection $\sigma$ along an arbitrary axis $\bm{\hat{z}}$, we can use eqs.~\eqref{simpleaeq} and \eqref{roteq} to express eq.~\eqref{diabdefeq} for $\bm{r}_0=0$ as
\begin{equation}
\ket{\zeta_i^{\eta,\sigma}(\bm{r}, 0)} = \sum_{\lambda,n} D_{\sigma,\lambda}^{s_i}(\varphi,\theta,\psi)^\ast A_{n,i}(r,0) \ket{\zeta_n^{\eta,\lambda}(\bm{r})}
\label{staterot}
\end{equation}
where $s_i$ is the spin of channel $i$. Plugging eq.~\eqref{staterot} into eq.~\eqref{potdefeq}, the diabatic potential matrix elements can be expressed as
\begin{equation}
V_{i,i^\prime}^{\eta,\sigma,\sigma^\prime}(\bm{r}) = \sum_\lambda D_{\sigma,\lambda}^{s_i}(\varphi,\theta,\psi) D_{\sigma^\prime,\lambda}^{s_{i^\prime}}(\varphi,\theta,\psi)^\ast G_{i,i^\prime}^{\eta,\lambda}(r),
\label{poteq}
\end{equation}
with
\[
G_{i,i^\prime}^{\eta,\lambda}(r) = \sum_n A_{n,i}(r,0)^\ast A_{n,i^\prime}(r,0) V_n^{\eta,\lambda}(r)
\]
the elements of the interaction matrices $\bm{G}^{\eta,\lambda}(r)$ given in eqs.~\eqref{g0mesmateq}-\eqref{upm1mesmateq}. Notice that eq.~\eqref{poteq} cannot be simply expressed as a matrix equation relating the $(\sigma,\sigma^\prime)$ block of the diabatic potential matrix, $\bm{V}^{\eta,\sigma,\sigma^\prime}(r)$, and the interaction matrices $\bm{G}^{\eta,\lambda}(r)$, because the spin quantum numbers $s_i$ and $s_{i^\prime}$ depend on $i$ and $i^\prime$.

The diabatic potential matrix $\bm{V}^\eta(\bm{r})$, which determines the motion of the heavy quarks through the master Schr\"odinger eq.~\eqref{schreq}, is built by a precise combination of the interaction matrices $\bm{G}^{\eta,\lambda}(r)$ for different values of $\lambda$. Note that the block $\sigma=\sigma^\prime=\lambda$ of the diabatic potential matrix coincides with $\bm{G}^{\eta,\lambda}(r)$ in the special case $\bm{\hat{r}}= \bm{\hat{z}}$:
\begin{equation}
\bm{V}^{\eta,\lambda,\lambda}(r \bm{\hat{z}}) = \bm{G}^{\eta,\lambda}(r).
\label{specialcaseq}
\end{equation}

It is instructive to verify that the introduction of the heavy-quark motion automatically reestablishes the full $O(3)\otimes C$ symmetry of QCD that was broken in the static approximation. As illustrated in appendix~\ref{spinorbap}, one can expand the heavy quark-antiquark state in eq.~\eqref{diabexpeq} in partial waves and observe that the diabatic potential matrix $\bm{V}^\eta(\bm{r})$ only connects partial waves with the same values of the total angular momentum quantum numbers $J$ and $M$. The Schr\"odinger eq.~\eqref{schreq} for wave functions that depend on the vector $\bm{r}$ gets transformed into a system of coupled radial Schr\"odinger equations,
\begin{equation}
\sum_{i^\prime,l^\prime} \biggl[\delta_{i,i^\prime} \delta_{l,l^\prime}\frac{1}{m_Q}\biggl(-\frac{\mathrm{d}^2\hphantom{r}}{\mathrm{d}r^2} + \frac{l(l+1)}{r^2}\biggr) + V_{i,i^\prime,l,l^\prime}^{\eta,J}(r) \biggr] u_{i^\prime,l^\prime}^{\eta,J}(r) = E u_{i,l}^{\eta,J}(r),
\label{radialeq}
\end{equation}
where $u_{i,l}^{\eta,J}(r)$ is the reduced radial wave function for the partial-wave channel $i$ with orbital momentum $l$, spin $s_i$, $CP$-parity $\eta$, and total angular momentum $J$, and $V_{i,i^\prime,l,l^\prime}^{\eta,J}(r)$ is the partial-wave potential
\begin{equation}
V_{i,i^\prime,l,l^\prime}^{\eta,J}(r) = \sqrt{(2l+1) (2l^\prime+1)}
\sum_\lambda
\begin{pmatrix}
s_i 		& l 		& J 		\\
\lambda	& 0		& -\lambda	\\
\end{pmatrix}
\begin{pmatrix}
s_{i^\prime} 	& l^{\prime} 	& J 		\\
\lambda	& 0		&  -\lambda	\\
\end{pmatrix}
G_{i,i^\prime}^{\eta,\lambda}(r)
\label{resulteq}
\end{equation}
with $\left(\begin{smallmatrix}j_1&j_2&j_3\\m_1&m_2&m_3\end{smallmatrix}\right)$ the Wigner 3-$j$ symbols. Notice that eq.~\eqref{resulteq} does not depend on the total angular momentum projection $M$, as required by the Wigner-Eckart theorem, so that the same Schr\"odinger eq.~\eqref{radialeq} (and therefore the same wave function) applies to all the $2J+1$ values of $M$. This suffices to show that the BO Hamiltonian respects the full rotational symmetry group $SO(3)$. Moreover, in appendix~\ref{symap} we show that the symmetry under a reflection $\mathcal{R}$ of $H_\textup{static}(\bm{r})$ ensures that the partial-wave potentials in eq.~\eqref{resulteq} conserve parity. This, combined with symmetry under the combined transformation $CP$, implies that $C$-parity is also conserved. We have thus verified that the relevant symmetry group of the BO Hamiltonian is $O(3)\otimes C$, so that the heavy quark-antiquark states calculated from it are naturally organized in $J^{PC}$ families.

In practice, plugging the interaction matrices in eqs.~\eqref{g0mesmateq}-\eqref{upm1mesmateq} into eq.~\eqref{resulteq} yields simple expressions for the partial-wave potentials in terms of the $Q\bar{Q}$ static energy $V_\mathcal{Q}(r)$, the $B^\ast$-$B$ mass difference $\Delta$, and the string-breaking transition rate $g(r)$. Specifically, if one neglects heavy-quark spin corrections of order $1/m_Q^2$ and above, the $Q\bar{Q}$ diagonal element of the interaction matrices $\bm{G}^{\eta,\lambda}(r)$ is the same regardless of the value of either $\eta$ or $\lambda$,
\[
G_{Q\bar{Q}(0),Q\bar{Q}(0)}^{u,0}(r) = G_{Q\bar{Q}(1),Q\bar{Q}(1)}^{g,0}(r) = G_{Q\bar{Q}(1),Q\bar{Q}(1)}^{g,\pm1}(r) = V_ {\mathcal{Q}}(r).
\]
Combining this with the orthogonality relations of the Wigner 3-$j$ symbols, the sum on the right side of eq.~\eqref{resulteq} gives a simple result for the $Q\bar{Q}$ potentials,
\begin{equation}
V_{Q\bar{Q}(0),Q\bar{Q}(0),l,l^\prime}^{u,J}(r) = V_{Q\bar{Q}(1),Q\bar{Q}(1),l,l^\prime}^{g,J}(r) = \delta_{l,l^\prime} V_\mathcal{Q}(r).
\label{simpleqqeq}
\end{equation}

Similarly, if one neglects light-meson exchange contributions and heavy-quark spin corrections of order $1/m_Q^2$ and above, the di-meson potentials are given by the corresponding threshold mass values relative to the spin-average $B^{(\ast)}\bar{B}^{(\ast)}$ threshold. One has
\[
V_{B\bar{B}(0),B\bar{B}(0),l,l^\prime}^{g,J}(r) = - \delta_{l,l^\prime}\frac{3}{2}\Delta
\]
for $B\bar{B}$,
\[
V_{B\bar{B}^\ast(1),B\bar{B}^\ast(1),l,l^\prime}^{g,J}(r) = V_{B\bar{B}^\ast(1),B\bar{B}^\ast(1),l,l^\prime}^{u,J}(r) = -\delta_{l,l^\prime} \frac{1}{2} \Delta
\]
for $B\bar{B}^\ast$, and
\[
V_{B^\ast\bar{B}^\ast(0),B^\ast\bar{B}^\ast(0),l,l^\prime}^{g,J}(r) = V_{B^\ast\bar{B}^\ast(1),B^\ast\bar{B}^\ast(1),l,l^\prime}^{u,J}(r) = V_{B^\ast\bar{B}^\ast(2),B^\ast\bar{B}^\ast(2),l,l^\prime}^{g,J}(r) = \delta_{l,l^\prime} \frac{1}{2} \Delta
\]
for $B^\ast\bar{B}^\ast$. Partial-wave potentials coupling different di-meson channels are zero.

Finally, under the reasonable assumption that HQSS remains a good approximation for the transition rates even when threshold spin splittings are included, any partial-wave potential coupling quark-antiquark and di-meson channels is given by a fraction of the string-breaking transition rate $g(r)$,
\begin{equation}
V_{Q\bar{Q}(s),B^{(\ast)}\bar{B}^{(\ast)}(s^\prime),l,l^\prime}^{\eta,J}(r) = \mathfrak{g}_{B^{(\ast)}\bar{B}^{(\ast)}}^{\eta,J,l,l^\prime,s,s^\prime} g(r)
\label{coefeq}
\end{equation}
with $\mathfrak{g}_{B^{(\ast)}\bar{B}^{(\ast)}}^{\eta,J,l,l^\prime,s,s^\prime}$ a scalar coefficient that can be calculated by inserting the interaction matrix elements from eqs.~\eqref{sg+eqs}-\eqref{su-eqs} into eq.~\eqref{resulteq}. For the sake of completeness, we list in appendix~\ref{coefap} all the nonvanishing values of $\mathfrak{g}_{B^{(\ast)}\bar{B}^{(\ast)}}^{\eta,J,l,l^\prime,s,s^\prime}$ with $J=0,1,2$.

It is worth emphasizing that, taking the coefficients $\mathfrak{g}_{B^{(\ast)}\bar{B}^{(\ast)}}^{\eta,J,l,l^\prime,s,s^\prime}$ from appendix~\ref{coefap}, the coupling potentials in eq.~\eqref{coefeq} for $S$-wave ($l=0$) $Q\bar{Q}$ coupling to $P$-wave ($l^\prime=1$) $B\bar{B}$, $B\bar{B}^{\ast g}$, $B^\ast\bar{B}^\ast(0)$, and $B^\ast\bar{B}^\ast(2)$ in $J^{PC}=1^{--}$ have relative ratios of
\[
-\frac{1}{2\sqrt{3}}:\frac{1}{\sqrt{3}}:-\frac{1}{6}:\frac{\sqrt{5}}{3},
\]
respectively. These ratios reproduce those in ref.~\cite{Vol12} for the amplitudes for production of $B\bar{B}$, $B^\ast\bar{B}+\text{c.c.}$, $(B^\ast\bar{B}^\ast)_{S=0}$, and $(B^\ast\bar{B}^\ast)_{S=2}$ in $e^+ e^-$ annihilation in the limit of exact HQSS (see eq.~(5) of ref.~\cite{Vol12}), up to sign differences due to phase conventions.

The coefficients $\mathfrak{g}_{B^{(\ast)}\bar{B}^{(\ast)}}^{\eta,J,l,l^\prime,s,s^\prime}$, fully constrained by HQSS and cylindrical symmetry, are always smaller than $1$ and differ considerably between one another; see appendix~\ref{coefap}. This finding is in stark contrast with the naive assumption $\mathfrak{g}_{B^{(\ast)}\bar{B}^{(\ast)}}^{\eta,J,l,l^\prime,s,s^\prime}=1$ for all quantum numbers made in the previous phenomenological analyses in refs.~\cite{Bru20,Bru21c,Bru22}. These analyses, which also use the diabatic representation of BO, did not incorporate the correct symmetries of the static approximation.

\subsection{Practical applications}

As a last remark, we illustrate how the diabatic BO approximation can be used in practical calculations of the quarkoniumlike spectrum.

Since $J$, $P$, and $C$ are all conserved quantum numbers, one can concentrate on a specific $J^{PC}$ configuration and consider only the corresponding orbital angular momentum channels when calculating the spectrum. From eq.~\eqref{radialeq}, one sees that the BO Hamiltonian in a particular $J^{PC}$ configuration can be represented as a matrix $\bm{H}^{J^{PC}}$ acting on the reduced radial wave functions. It can be expressed as
\[
\bm{H}^{J^{PC}} = - \frac{1}{m_Q} \frac{\mathrm{d}^2\hphantom{r}}{\mathrm{d}r^2} + \bm{V}^{J^{PC}}(r),
\]
where the entries of the radial potential matrix $\bm{V}^{J^{PC}}(r)$ have the form
\begin{equation}
V_{i,i^\prime,l,l^\prime}^{J^{PC}}(r) = V_{i,i^\prime,l,l^\prime}^{\eta,J}(r) + \delta_{i,i^\prime} \delta_{l,l^\prime} \frac{l(l+1)}{m_Q r^2}.
\label{radpoteq}
\end{equation}
Note that $\eta$, which determines the relevant quark-antiquark and di-meson channels, corresponds to the product of the given $P$ and $C$, and that the values of $l$ and $l^\prime$ are constrained by conservation of total angular momentum and parity. The eigenvalues and eigenvectors of $\bm{H}^{J^{PC}}$ can be calculated numerically. From them, the masses of quarkoniumlike bound states and the open-flavor di-meson scattering amplitudes can be readily obtained; see ref.~\cite{Bru21c}.

It is instructive to spell out in full the radial potential matrix $\bm{V}^{J^{PC}}(r)$ in eq.~\eqref{radpoteq} in the simplest possible case, $J^{PC}=0^{-+}$, corresponding to $Q\bar{Q}$ with spin $0$ and orbital angular momentum $0$. There are only 3 partial-wave channels: $Q\bar{Q}(0)$ with $l=0$, $B\bar{B}^{\ast u}(1)$ with $l=1$, and $B^\ast\bar{B}^\ast(1)$ with $l=1$. The corresponding radial potential is the $3\times3$ matrix
\begin{equation}
\bm{V}^{0^{-+}}(r) =
\begin{pmatrix}
V_{\mathcal{Q}}(r)		& \frac{1}{\sqrt{2}}g(r)	& \frac{1}{\sqrt{2}}g(r)	\\
\frac{1}{\sqrt{2}}g(r)	& -\frac{1}{2}\Delta + \frac{2}{m_Q r^2}	& 0					\\
\frac{1}{\sqrt{2}}g(r)	& 0						& \frac{1}{2}\Delta + \frac{2}{m_Q r^2}	\\
\end{pmatrix}.
\label{0-+eq}
\end{equation}

As a further consistency check, it is useful to verify that for $\Delta\to0$ the spectrum of $\bm{H}^{0^{-+}}$ is independent of the $Q\bar{Q}$ spin, as expected from restoration of HQSS in this limit. Taking the limit $\Delta\to0$ in eq.~\eqref{0-+eq} yields
\[
\lim_{\Delta\to0}\bm{V}^{0^{-+}}(r) =
\begin{pmatrix}
V_{\mathcal{Q}}(r)		& \frac{1}{\sqrt{2}}g(r)	& \frac{1}{\sqrt{2}}g(r)	\\
\frac{1}{\sqrt{2}}g(r)	& \frac{2}{m_Q r^2}	& 0					\\
\frac{1}{\sqrt{2}}g(r)	& 0						&  \frac{2}{m_Q r^2}	\\
\end{pmatrix}.
\]
The second and third diagonal elements of the potential matrix are now identical. This allows one to perform a unitary transformation that turns the potential matrix into
\[
\lim_{\Delta\to0}\bm{V}^{0^{-+}}(r) =
\begin{pmatrix}
\bm{V}_\textup{HQSS}^{0^{-+}}(r)	& 0					\\
0						& \frac{2}{m_Q r^2}		\\
\end{pmatrix}.
\]
The last channel is just a trivial free-wave channel decoupled from the others. The other two channels form the $2\times2$ potential matrix with HQSS,
\begin{equation}
\bm{V}_\textup{HQSS}^{0^{-+}}(r) =
\begin{pmatrix}
V_{\mathcal{Q}}(r)		& g(r)	\\
g(r)	& \frac{2}{m_Q r^2}	\\
\end{pmatrix}.
\label{hqss0-+eq}
\end{equation}

Let us now take the $J^{PC}=1^{--}$ case, corresponding to $Q\bar{Q}$ with spin $1$ and angular momentum either $0$ or $2$. There are 7 partial-wave channels: $Q\bar{Q}(1)$ with $l=0,2$, $B\bar{B}(0)$ with $l=1$, $B\bar{B}^{\ast g}(1)$ with $l=1$, $B^\ast\bar{B}^\ast(0)$ with $l=1$, and $B^\ast\bar{B}^\ast(2)$ with $l=1,3$. The corresponding radial potential is the $7\times7$ matrix
\begin{subequations}
\begin{equation}
\bm{V}^{1^{--}}(r) =
\begin{pmatrix}
\bm{V}_\mathcal{Q}^{1^{--}}(r)		& \bm{V}^{1^{--}}_\textup{mix}(r)	\\
\bm{V}^{1^{--}}_\textup{mix}(r)^\dag	& \bm{V}_\mathcal{B}^{1^{--}}(r)	\\
\end{pmatrix}
\end{equation}
with
\begin{equation}
\bm{V}_\mathcal{Q}^{1^{--}}(r) = V_{\mathcal{Q}}(r) + \diag(0, 6) \frac{1}{m_Q r^2}
\end{equation}
the $2\times2$ quark-antiquark potential sub-matrix,
\begin{equation}
\bm{V}_\mathcal{B}^{1^{--}}(r) = \diag(-3,-1,1,1,1) \frac{\Delta}{2} + \diag(2, 2, 2, 2, 12) \frac{1}{m_Q r^2}
\end{equation}
the $5\times5$ di-meson potential sub-matrix, and
\begin{equation}
\bm{V}^{1^{--}}_\textup{mix}(r) =
\begin{pmatrix}
-\frac{1}{2\sqrt{3}}	& \frac{1}{\sqrt{3}}	& -\frac{1}{6}			& \frac{\sqrt{5}}{3}		& 0 						\\
\frac{1}{\sqrt{6}}	& \frac{1}{\sqrt{6}} 	& \frac{1}{3\sqrt{2}}	& -\frac{1}{3\sqrt{10}}	& \sqrt{\frac{3}{5}}			\\
\end{pmatrix} g(r)
\end{equation}
the $2\times5$ mixing potential sub-matrix.
\label{1--eq}
\end{subequations}

In the limit $\Delta\to0$, there is a unitary transformation that turns the $7\times7$ potential matrix in eqs.~\eqref{1--eq} into
\[
\lim_{\Delta\to0}\bm{V}^{1^{--}}(r) =
\begin{pmatrix}
\bm{V}_\textup{HQSS}^{1^{--}}(r)	& 0					& 0					\\
0								& \frac{2}{m_Q r^2}	& 0					\\
0								& 0					& \frac{2}{m_Q r^2}	\\
\end{pmatrix}.
\]
The last two channels are just trivial free-wave channels decoupled from the others. The other five channels form the $5\times5$ potential matrix with HQSS,
\begin{equation}
\bm{V}_\textup{HQSS}^{1^{--}}(r) =
\begin{pmatrix}
\bm{V}_\textup{HQSS}^{0^{-+}}(r)		& 0									\\
0									& \bm{V}_\textup{HQSS}^{2^{-+}}(r)	\\
\end{pmatrix}
\label{hqss1--eq}
\end{equation}
with $\bm{V}_\textup{HQSS}^{0^{-+}}(r)$ the $2\times2$ potential matrix in eq.~\eqref{hqss0-+eq} and
\begin{equation}
\bm{V}_\textup{HQSS}^{2^{-+}}(r) =
\begin{pmatrix}
V_{\mathcal{Q}}(r) + \frac{6}{m_Q r^2}	& \sqrt{\frac{2}{5}}g(r)	& \sqrt{\frac{3}{5}}g(r)	\\
\sqrt{\frac{2}{5}}g(r)			& \frac{2}{m_Q r^2}	& 0					\\
\sqrt{\frac{3}{5}}g(r)			& 0					& \frac{12}{m_Q r^2}	\\
\end{pmatrix}
\label{hqss2-+eq}
\end{equation}
the $3\times3$ potential matrix for $J^{PC}=2^{-+}$ with HQSS.

It is clear from eqs.~\eqref{hqss0-+eq}, \eqref{hqss1--eq}, and \eqref{hqss2-+eq} that, in the limit $\Delta\to0$, the $Q\bar{Q}$ $S$-wave ($l=0$) and $D$-wave ($l=2$) channels in the $J^{PC}=1^{--}$ configuration are completely decoupled from each other. This also shows that, in this HQSS limit, the spectrum of quarkoniumlike states with $Q\bar{Q}$ in $S$-wave is the same regardless of whether the $Q\bar{Q}$ spin is $0$ or $1$ ($J^{PC}=0^{-+}$ or $1^{--}$), as expected.

These results should be compared with the BO calculation using lattice-QCD BO potentials with string breaking in ref.~\cite{Bic20}, where the spectrum of $\Upsilon$ and $\eta_b$ states is also calculated from a coupled-channel radial Schr\"odinger equation for quark-antiquark and di-meson channels. One can notice that the radial potential matrix of this reference, which may be easily read from the left side of eq.~(50) in ref.~\cite{Bic20}, is identical to our $\bm{V}_\textup{HQSS}^{0^{-+}}(r)$ in eq.~\eqref{hqss0-+eq} by substituting $m_Q \to 2 \mu_M$, $V_\mathcal{Q}(r) \to V_{\bar{Q}Q}(r)$, $g(r) \to V_\textup{mix}(r)$, and noticing that $V_{\bar{M}M,\parallel}(r)$ in eq.~(50) of ref.~\cite{Bic20} is actually $0$ (see eq.~(59) in ref.~\cite{Bic20}).

Therefore, our results reproduce the Schr\"odinger equation used in ref.~\cite{Bic20} in the HQSS limit $\Delta\to0$. When threshold spin splittings $\Delta>0$ are included, however, we observe breaking of HQSS through the difference between the radial potential matrices in the cases of $Q\bar{Q}$ spin $0$ and $1$, $\bm{V}^{0^{-+}}(r)$ and $\bm{V}^{1^{--}}(r)$. This HQSS-breaking effect is straightforwardly incorporated into the lattice-QCD potentials with string breaking using our eqs.~\eqref{0-+eq} and \eqref{1--eq} for the specific $0^{-+}$ and $1^{--}$ cases, or eqs.~\eqref{resulteq} and \eqref{radpoteq} for any $J^{PC}$ in general. The same cannot be said of the results in refs.~\cite{Bic20,Bic21}, for which no clear prescription to go beyond the HQSS approximation is provided.

Notice that, since the heavy-quark kinetic energy and the $B^\ast$-$B$ mass splitting are both first-order corrections in $1/m_Q$ to the static approximation, there is no formal justification for including the former without the latter in a BO approximation. From a phenomenological point of view, this HQSS-breaking effect is kinematically suppressed for quarkonium states with masses well below the $B\bar{B}$ threshold, where di-meson channels hardly play any role, but it is relevant in all other cases. This is especially important for exotic candidates, whose masses often lie close to some open-flavor meson-pair threshold.

For these reasons, the diabatic BO approximation studied here should be regarded as a necessary improvement of the scheme used in refs.~\cite{Bic20,Bic21}, which allows one to perform detailed studies for energies close and above threshold.

\section{Overview}
\label{oversec}

We have revisited the static approximation for a heavy quark-antiquark pair. In addition to the $Q\bar{Q}$ and $B\bar{B}$ sources, we have also included $B\bar{B}^\ast$ and $B^\ast\bar{B}^\ast$ sources. We have shown that the transition rates between quark-antiquark and di-meson configurations are given by fractions of the string-breaking transition rate, with coefficients completely determined from Dirac algebra.

We have then introduced the mass splitting between a heavy-light pseudoscalar meson $B$ and its vector partner $B^\ast$. We have illustrated the pattern of HQSS breaking deriving from string breaking with threshold spin splittings. We have further shown that this symmetry breaking is easily accounted for in the interaction matrix between quark-antiquark and di-meson configurations.

Finally, we have built a systematic treatment of the BO approximation at first order in $1/m_Q$ in the diabatic representation. We have shown that the dynamics of the heavy quark-antiquark system, including the heavy quark-antiquark interaction, string breaking, and HQSS breaking, is governed by the diabatic potential matrix, which is completely determined from an interaction matrix accessible in lattice QCD. We have also shown, without any \textit{ad hoc} assumption, that this construction of the BO approximation leads to a natural restoration of the $O(3)\otimes C$ symmetry of QCD from the cylindrical symmetry $D_{\infty h}\otimes CP$ of the static approximation.

This result provides an essential stepping stone towards precise, first-principles studies of quarkoniumlike mesons and open-flavor di-meson scattering with string breaking. It is worth highlighting that the inclusion of threshold spin splittings is particularly relevant to the understanding of any heavy quark-antiquark state in which meson pairs play a significant role, and to the understanding of near-threshold states like $X(3872)$ in particular.

For the charmoniumlike sector, it may be necessary to consider even the small $D^+$-$D^0$ and $D^{\ast+}$-$D^{\ast0}$ mass differences. For instance, they may be instrumental in explaining the isospin-violating branching fractions of $X(3872)$ \cite{dAm10,Abl19}. Note that such isospin mass splittings can be easily incorporated into the diabatic BO approximation; see ref.~\cite{Bru22}.

For practical applications in the future, it may be necessary to include also other relevant effects, like heavy-quark spin corrections to the quark-antiquark potential of order $1/m_Q^2$ and light-meson exchange interactions between two open-flavor mesons. The former are well known \cite{Eic81,Pin01,Kom07} and can be simply added to the $Q\bar{Q}$ partial-wave potentials. As for the latter, we strongly encourage a theoretical effort for their incorporation in the diabatic BO approximation.

\acknowledgments{
I wish to express my gratitude to E. Braaten for his indispensable help in developing this project. I acknowledge valuable discussions with P. Gonz\'alez. This research was supported by the U.S. Department of Energy under Grant No.~\texttt{DE-SC0011726}.
}

\appendix

\section{Reflection symmetry}
\label{reflap}

Let us consider a static quark-antiquark source producing a light BO state with quantum numbers $J_\textup{light}$ and $\epsilon_\textup{light}$ (defined by the $\Lambda=0$ member of the multiplet), where the total spin $\bm{S}_{Q\bar{Q}}$ of the heavy quarks combines with the light QCD angular momentum $\bm{J}_\textup{light}$ to give definite values of the static spin $s$ and its projection $\lambda$ along the quark-antiquark separation vector $\bm{r}$. To calculate how this state transforms under a reflection $\mathcal{R}$ through a plane containing $\bm{r}$, one may decompose such transformation as:
\begin{itemize}
\item parity inversion through the midpoint of $\bm{r}$, and
\item rotation of an angle $\pi$ around an axis perpendicular to $\bm{r}$ and passing through its midpoint.
\end{itemize}

Let us work in the center-of-mass frame, so that the origin coincides with the midpoint of $\bm{r}$, and identify the arbitrary axis $\bm{\hat{z}}$ with $\bm{\hat{r}}$, for simplicity. One can then express a reflection $\mathcal{R}$ as
\begin{equation}
\mathcal{R} = e^{-i\pi J_y}  P
\label{refleq}
\end{equation}
with $P$ the operator for parity inversion through the origin. 

The rotation $e^{-i\pi J_y}$ swaps places between the static sources and flips the value of the static spin projection, $\lambda\to-\lambda$, while producing a multiplicative phase of $(-1)^{-s}$.

Parity inversion $P$ swaps back places between the static sources and produces two more multiplicative phases. One is the intrinsic parity $\mathfrak{p}$ of the source, given by the product of the intrinsic parities of the two individual sources. The other multiplicative phase corresponds to the intrinsic parity $P_\textup{light}$ of the light BO state. This can be calculated by applying $\mathcal{R}$ to the light BO state in the limit $r\to 0$, where it becomes an eigenstate of $P$ with eigenvalue $P_\textup{light}$. From eq.~\eqref{refleq} one obtains $\epsilon_\textup{light} = P_\textup{light}  (-1)^{-J_\textup{light}}$, and thus $P_\textup{light} = \epsilon_\textup{light} (-1)^{J_\textup{light}}$.

All things considered, the overall effect of $\mathcal{R}$ on a static BO state formed by the light QCD fields and the spins of the heavy quarks amounts to the inversion of $\lambda$ and multiplication by a sign
\[
\epsilon = P_\textup{light} \mathfrak{p} (-1)^{-s} = \epsilon_\textup{light}\mathfrak{p}(-1)^{J_\textup{light}-s}.
\]
Therefore, a static BO state with static spin projection $\lambda=0$ is an eigenstate of $\mathcal{R}$ with eigenvalue $\epsilon$.

\section{Conservation of total angular momentum}
\label{spinorbap}

To verify that total angular momentum is conserved when the orbital angular momentum of the sources is reintroduced, let us expand eq.~\eqref{diabexpeq} in partial waves as
\[
\ket{\Psi} =  \sum_{\eta,J,M,i,l} \int \mathrm{d}r \, u_{i,l}^{\eta,J,M}(r) \ket{r} \ket{\zeta_{i,l}^{\eta,J,M}(r,0)}
\]
with $\ket{\zeta_{i,l}^{\eta,J,M}(r,0)}$ the partial-wave channel $i$ with orbital momentum $l$, spin $s_i$, $CP$-parity $\eta$, total angular momentum $J$, and projection $M$, and $u_{i,l}^{\eta,J,M}(r)$ the corresponding reduced radial wave function.

In this expansion, the Schr\"odinger eq.~\eqref{schreq} translates into a system of coupled radial equations for the reduced radial wave functions $u_{i,l}^{\eta,J,M}(r)$ where the partial-wave potentials are given by
\begin{multline*}
V_{i,i^\prime,l,l^\prime}^{\eta,J,J^\prime,M,M^\prime}(r) = \\
\sum_{m,\sigma} \sum_{m^\prime,\sigma^\prime} \cgl{J}{M}{l}{m}{s_i}{\sigma} \cgr{l^\prime}{m^\prime}{s_{i^\prime}}{\sigma^\prime}{J^\prime}{M^\prime}
\int_0^{2\pi} \mathrm{d}\varphi \int_{-1}^{+1} \mathrm{d}\cos\theta\, Y_l^m(\theta,\varphi)^\ast  Y_{l^\prime}^{m^\prime}(\theta,\varphi) V_{i,i^\prime}^{\eta,\sigma,\sigma^\prime}(\bm{r}),
\end{multline*}
with $Y_l^m(\theta,\varphi)$ the spherical harmonics and $\smallcg{j_1}{m_1}{j_2}{m_2}{j_3}{m_3}$ the Clebsch-Gordan coefficients. Inserting eq.~\eqref{poteq} into the above yields
\begin{multline*}
V_{i,i^\prime,l,l^\prime}^{\eta,J,J^\prime,M,M^\prime}(r) = \sum_\lambda G_{i,i^\prime}^{\eta,\lambda}(r) \int_0^{2\pi} \mathrm{d}\varphi \int_{-1}^{+1} \mathrm{d}\cos\theta \\
\sum_{m,\sigma} D_{\sigma,\lambda}^{s_i}(\varphi,\theta,\psi) Y_l^m(\theta,\varphi)^\ast \cgl{J}{M}{l}{m}{s_i}{\sigma} \sum_{m^\prime,\sigma^\prime} D_{\sigma^\prime,\lambda}^{s_{i^\prime}}(\varphi,\theta,\psi)^\ast Y_{l^\prime}^{m^\prime}(\theta,\varphi) \cgr{l^\prime}{m^\prime}{s_{i^\prime}}{\sigma^\prime}{J^\prime}{M^\prime}.
\end{multline*}

The sum over $m$, $m^\prime$, $\sigma$, and $\sigma^\prime$ can be evaluated using the expression of the spherical harmonics in terms of Wigner $D$-matrix elements,
\[
Y_l^m(\theta,\varphi) = \sqrt{\frac{2l+1}{4\pi}} D_{m,0}^l(\varphi,\theta,\psi)^\ast,
\]
the expansion property of the Wigner $D$-matrices,
\[
D_{a,b}^{c}(\varphi,\theta,\psi) D_{a^\prime,b^\prime}^{c^\prime}(\varphi,\theta,\psi) = \sum_d D_{a + a^\prime,b+b^\prime}^d (\varphi,\theta,\psi) \cgr{c}{a}{c^\prime}{a^\prime}{d}{a+a^\prime} \cgl{d}{b+b^\prime}{c}{b}{c^\prime}{b^\prime},
\]
and the orthogonality relation of the Clebsch-Gordan coefficients. This gives
\[
\sum_{m,\sigma} D_{\sigma,\lambda}^{s_i}(\varphi,\theta,\psi) Y_l^m(\theta,\varphi)^\ast \cgl{J}{M}{l}{m}{s_i}{\sigma} = \sqrt{\frac{2l+1}{4\pi}} D_{M,\lambda}^J(\varphi,\theta,\psi) \cgl{J}{\lambda}{l}{0}{s_i}{\lambda}
\]
and therefore
\begin{multline*}
V_{i,i^\prime,l,l^\prime}^{\eta,J,J^\prime,M,M^\prime}(r) = \frac{\sqrt{(2l+1) (2l^\prime+1)}}{4\pi} \sum_\lambda G_{i,i^\prime}^{\eta,\lambda}(r) \cgl{J}{\lambda}{l}{0}{s_i}{\lambda} \cgr{l^\prime}{0}{s_{i^\prime}}{\lambda}{J^\prime}{\lambda} \\
\int_0^{2\pi} \mathrm{d}\varphi \int_0^\pi \mathrm{d}\theta \, \sin\theta D_{M,\lambda}^J(\varphi,\theta,\psi) D_{M^\prime,\lambda}^{J^\prime}(\varphi,\theta,\psi)^\ast.
\end{multline*}

Finally, the angular integrals can be evaluated using the orthogonality relation of the Wigner $D$-matrices. One gets
\[
\int_0^{2\pi} \mathrm{d}\varphi \int_0^\pi \mathrm{d}\theta \, \sin\theta D_{M,\lambda}^J(\varphi,\theta,\psi) D_{M^\prime,\lambda}^{J^\prime}(\varphi,\theta,\psi)^\ast = \frac{4\pi}{2J+1} \delta_{J,J^\prime} \delta_{M,M^\prime}
\]
and thus
\[
V_{i,i^\prime,l,l^\prime}^{\eta,J,J^\prime,M,M^\prime}(r) = \delta_{J,J^\prime} \delta_{M,M^\prime} V_{i,i^\prime,l,l^\prime}^{\eta,J}(r)
\]
with
\[
V_{i,i^\prime,l,l^\prime}^{\eta,J}(r) = \frac{\sqrt{(2l+1)(2l^\prime+1)}}{2 J + 1} \sum_\lambda G_{i,i^\prime}^{\eta,\lambda}(r) \cgl{J}{\lambda}{l}{0}{s_i}{\lambda} \cgr{l^\prime}{0}{s_{i^\prime}}{\lambda}{J}{\lambda}.
\]
In eq.~\eqref{resulteq}, we have expressed the Clebsh-Gordan coefficients in terms of Wigner 3-$j$ symbols using
\[
\cgr{j_1}{m_1}{j_2}{m_2}{j_3}{m_3} = (-1)^{2 j_2+j_3-m_3} \sqrt{2j_3 + 1}
\begin{pmatrix}
j_2	& j_1	& j_3		\\
m_2	& m_1	& -m_3	\\
\end{pmatrix}.
\]

\section{Conservation of parity}
\label{symap}

To prove that the partial-wave potentials of eq.~\eqref{resulteq} conserve parity, let us begin by showing that symmetry under a reflection $\mathcal{R}$ through a plane containing the heavy sources implies that each entry $G_{i,i^\prime}^{\eta,\lambda}(r)$ of the interaction matrices $\bm{G}^{\eta,\lambda}(r)$ is either symmetric or antisymmetric in $\lambda$. Combining together eqs.~\eqref{potdefeq} and \eqref{specialcaseq}, the entries of $\bm{G}^{\eta,\lambda}(r)$ can be expressed as
\[
G_{i,i^\prime}^{\eta,\lambda}(r) = \braket{\zeta_i^{\eta,\lambda}(r \bm{\hat{z}},0) | H_\textup{static}(r \bm{\hat{z}}) | \zeta_{i^\prime}^{\eta,\lambda}(r \bm{\hat{z}},0)}.
\]
Then, symmetry under a reflection $\mathcal{R}$, $[\mathcal{R}, H_\textup{static}(r \bm{\hat{z}})] = 0$, implies
\begin{equation}
G_{i,i^\prime}^{\eta,\lambda}(r) = \braket{\zeta_i^{\eta,\lambda}(r \bm{\hat{z}},0) | \mathcal{R}^\dagger H_\textup{static}(r \bm{\hat{z}}) \mathcal{R} | \zeta_{i^\prime}^{\eta,\lambda}(r \bm{\hat{z}},0)}.
\label{refleq1}
\end{equation}
Note that the right side of eq.~\eqref{refleq1} is evaluated for $\bm{\hat{r}}=\bm{\hat{z}}$, so the reflection operator $\mathcal{R}$ can be written as in eq.~\eqref{refleq}. The action of $\mathcal{R}$ on the diabatic states $\ket{\zeta_i^{\eta,\lambda}(r \bm{\hat{z}},0)}$ gives
\begin{equation}
\mathcal{R} \ket{\zeta_{i}^{\eta,\lambda}(r \bm{\hat{z}},0)} = \mathfrak{p}_i (-1)^{-s_i} \ket{\zeta_{i}^{\eta,-\lambda}(r \bm{\hat{z}},0)},
\label{refleq2}
\end{equation}
with $\mathfrak{p}_{Q\bar{Q}} = -1$ or $\mathfrak{p}_{B^{(\ast)}\bar{B}^{(\ast)}} = +1$ the intrinsic parity of the source. Plugging eq.~\eqref{refleq2} back into eq.~\eqref{refleq1} yields
\begin{equation}
G_{i,i^\prime}^{\eta,\lambda}(r) = \mathfrak{p}_i \mathfrak{p}_{i^\prime} (-1)^{s_i - s_{i^\prime}} G_{i,i^\prime}^{\eta,-\lambda}(r).
\label{refleq3}
\end{equation}
Thus, each $G_{i,i^\prime}^{\eta,\lambda}(r)$ is either symmetric or antisymmetric in $\lambda$.

Now consider the sum on the right side of eq.~\eqref{resulteq}. Per the symmetry properties of the Wigner 3-$j$ symbols, one has
\begin{multline}
\sum_\lambda
\begin{pmatrix}
s_i 		& l 		& J 		\\
\lambda	& 0		& -\lambda	\\
\end{pmatrix}
\begin{pmatrix}
s_{i^\prime} 	& l^{\prime} 	& J 		\\
\lambda	& 0		&  -\lambda	\\
\end{pmatrix}
G_{i,i^\prime}^{\eta,\lambda}(r) =\\
(-1)^{s_i+s_{i^\prime} + l + l^\prime+2J}
\sum_\lambda
\begin{pmatrix}
s_i 		& l 		& J 		\\
-\lambda	& 0		& \lambda	\\
\end{pmatrix}
\begin{pmatrix}
s_{i^\prime} 	& l^{\prime} 	& J 		\\
-\lambda	& 0		&  \lambda	\\
\end{pmatrix}
G_{i,i^\prime}^{\eta,\lambda}(r).
\label{refleq4}
\end{multline}
Inserting eq.~\eqref{refleq3} on the right side of eq.~\eqref{refleq4} and subtracting it from the left side yields
\[
\Bigl(1 - \mathfrak{p}_i \mathfrak{p}_{i^\prime} (-1)^{l+l^\prime}\Bigr)
\sum_\lambda
\begin{pmatrix}
s_i 		& l 		& J 		\\
\lambda	& 0		& -\lambda	\\
\end{pmatrix}
\begin{pmatrix}
s_{i^\prime} 	& l^{\prime} 	& J 		\\
\lambda	& 0		&  -\lambda	\\
\end{pmatrix}
G_{i,i^\prime}^{\eta,\lambda}(r)
= 0.
\]
For the sum to be nonzero, the prefactor must vanish. This shows that the partial-wave potential in eq.~\eqref{resulteq} can be different from zero only if the partial-wave channels satisfy
\[
\mathfrak{p}_i  (-1)^l = \mathfrak{p}_{i^\prime} (-1)^{l^\prime},
\]
that is, if they have the same parity.

This proof can be easily extended to the general case of light BO quantum numbers other than the $\Sigma_g^+$ case discussed here. The only difference is that eq.~\eqref{refleq2} has to be modified to take into account also the intrinsic parity of the light BO state, $P_\textup{light} = \epsilon_\textup{light} (-1)^{J_\textup{light}}$ (see appendix~\ref{reflap}). Note that in the general case parity is given by
\[
P = P_\textup{light} \mathfrak{p} (-1)^l = \epsilon_\textup{light}\mathfrak{p}(-1)^{l + J_\textup{light}}.
\]

\section{Partial-wave coupling coefficients}
\label{coefap}

In this appendix, we list some values of the numerical coefficients $\mathfrak{g}_{B^{(\ast)}\bar{B}^{(\ast)}}^{\eta,J,l,l^\prime,s,s^\prime}$ entering the partial-wave coupling potentials in eq.~\eqref{coefeq} between $Q\bar{Q}(s)$ with orbital angular momentum $l$ and $B^{(\ast)}\bar{B}^{(\ast)}(s^\prime)$ with orbital angular momentum $l^\prime$. The coefficients for $\eta=g$ and $J=0,1,2$ are given in Table~\ref{gtable}. The coefficients for $\eta=u$ and $J=0,1,2$ are given in Table~\ref{utable}.

\begin{table}
\caption{\label{gtable}Partial-wave coupling coefficients $\mathfrak{g}_{B^{(\ast)}\bar{B}^{(\ast)}}^{\eta,J,l,l^\prime,s,s^\prime}$ in $\eta=g$ ($CP=+$) and $J=0,1,2$ between $Q\bar{Q}(1)$ orbital angular momentum channels $l$ and di-meson orbital angular momentum channels $l^\prime$. Notice that, for each $J^{PC}$ and each $l$ within it, the sum of the squares of the coefficients over $l^\prime$ and the 4 di-meson channels is 1.}
\begin{center}
\begin{tabular}{cccccc}
\toprule
$J^{PC}$	& $(l,l^\prime)$	& $B\bar{B}(0)$					& $B\bar{B}^\ast(1)$		& $B^\ast\bar{B}^\ast(0)$	& $B^\ast\bar{B}^\ast(2)$	\\
\midrule
$0^{++}$	& $(1,0)$		&  $\frac{1}{2}$				& 0					& $\frac{1}{2\sqrt{3}}$	& 0					\\
		& $(1,2)$		& 0							& 0					& 0					& $\sqrt{\frac{2}{3}}$	\\[.2em]
\midrule
$1^{++}$	& $(1,0)$		& 0							& $\frac{1}{\sqrt{3}}$	& 0					& 0					\\
		& $(1,2)$		& 0							& $\frac{1}{\sqrt{6}}$	& 0					& $\frac{1}{\sqrt{2}}$	\\[.2em]
\midrule
$1^{--}$	& $(0,1)$		& $-\frac{1}{2\sqrt{3}}$			& $\frac{1}{\sqrt{3}}$	& $-\frac{1}{6}$		& $\frac{\sqrt{5}}{3}$	\\
		& $(2,1)$		& $\frac{1}{\sqrt{6}}$			& $\frac{1}{\sqrt{6}}$	& $\frac{1}{3\sqrt{2}}$	& $-\frac{1}{3\sqrt{10}}$	\\
		& $(2,3)$		& 0							& 0					& 0					& $\sqrt{\frac{3}{5}}$	\\[.2em]
\midrule
$2^{++}$	& $(1,0)$		& 0							& 0					& 0					& $-\frac{1}{\sqrt{3}}$	\\
		& $(1,2)$		& $-\frac{1}{\sqrt{10}}$			& $\sqrt{\frac{3}{10}}$	& $-\frac{1}{\sqrt{30}}$	& $\sqrt{\frac{7}{30}}$	\\
		& $(3,2)$		& $\frac{1}{2}\sqrt{\frac{3}{5}}$	& $\frac{1}{\sqrt{5}}$	& $\frac{1}{2\sqrt{5}}$	& $-\frac{1}{\sqrt{35}}$	\\
		& $(3,4)$		& 0							& 0					& 0					& $\frac{2}{\sqrt{7}}$	\\[.2em]
\midrule
$2^{--}$	& $(2,1)$		& 0							& $\sqrt{\frac{3}{10}}$	& 0					& $-\frac{1}{\sqrt{10}}$	\\
		& $(2,3)$		& 0							& $\frac{1}{\sqrt{5}}$	& 0					& $\sqrt{\frac{2}{5}}$	\\
\bottomrule
\end{tabular}
\end{center}
\end{table}

\begin{table}
\caption{\label{utable}Partial-wave coupling coefficients $\mathfrak{g}_{B^{(\ast)}\bar{B}^{(\ast)}}^{\eta,J,l,l^\prime,s,s^\prime}$ in $\eta=u$ ($CP=-$) and $J=0,1,2$ between $Q\bar{Q}(0)$ orbital angular momentum channels $l$ and di-meson orbital angular momentum channels $l^\prime$. Notice that, for each $J^{PC}$ and each $l$ within it, the sum of the squares of the coefficients over $l^\prime$ and the 2 di-meson channels is 1.}
\begin{center}
\begin{tabular}{cccc}
\toprule
$J^{PC}$	&$(l,l^\prime)$	& $B\bar{B}^\ast(1)$		& $B^\ast\bar{B}^\ast(1)$	\\
\midrule
$0^{-+}$	& $(0,1)$		& $\frac{1}{\sqrt{2}}$	& $\frac{1}{\sqrt{2}}$	\\[.2em]
\midrule
$1^{+-}$	& $(1,0)$		& $-\frac{1}{\sqrt{6}}$	& $-\frac{1}{\sqrt{6}}$	\\
		& $(1,2)$		& $\frac{1}{\sqrt{3}}$	& $\frac{1}{\sqrt{3}}$	\\[.2em]
\midrule
$2^{-+}$	& $(2,1)$		& $-\frac{1}{\sqrt{5}}$	& $-\frac{1}{\sqrt{5}}$	\\
		& $(2,3)$		&$\sqrt{\frac{3}{10}}$	& $\sqrt{\frac{3}{10}}$	\\
\bottomrule
\end{tabular}
\end{center}
\end{table}

\bibliography{breakingbib}

\end{document}